\documentclass[preprint]{elsarticle}

\usepackage{bm,amsmath,amsfonts,amssymb}
\usepackage{hyperref}       
\usepackage{graphicx} 
\usepackage{nicefrac}       
\usepackage{url}            
\usepackage{doi}
\usepackage{orcidlink}
\usepackage{verbatim}

\begin{document}

\begin{frontmatter}
\title{Converting sWeights to Probabilities with Density Ratios}

\author[1]{D.I.~Glazier\,\orcidlink{0000-0002-8929-6332}}
\author[2]{R.~Tyson\,\orcidlink{0000-0002-0635-4198}}
\affiliation[1]{organization={SUPA, School of Physics and Astronomy, University of Glasgow}, 
                city={Glasgow}, 
                postcode={G12 8QQ}, 
                country={United Kingdom}}
\affiliation[2]{organization={Thomas Jefferson National Accelerator Facility}, 
                city={Newport News}, 
                postcode={23606},
                country={USA}}

\begin{abstract}
The use of machine learning approaches continues to have many benefits in experimental nuclear and particle physics. One common issue is generating training data which is sufficiently realistic to give reliable results. Here we advocate using real experimental data as the source of training data and demonstrate how one might subtract background contributions through the use of probabilistic weights which can be readily applied to training data.
 The \emph{sPlot} formalism is a common tool used to isolate distributions from different sources. However, the negative \emph{sWeights} produced by the \emph{sPlot} technique can cause training problems and poor predictive power. This article demonstrates how density ratio estimation can be applied to convert \emph{sWeights} to event probabilities, which we call \emph{drWeights}. The \emph{drWeights} can then be applied to produce the distributions of interest and are consistent with direct use of the \emph{sWeights}. This article will also show how decision trees are particularly well suited to convert \emph{sWeights}, with the benefit of fast prediction rates and adaptability to aspects of experimental data such as the data sample size and proportions of different event sources. We also show that a density ratio product approach in which the initial \emph{drWeights} are reweighted by an additional converter gives substantially better results.

\end{abstract}

\end{frontmatter}

\section{Introduction} \label{sect_intro}
\noindent
A significant complication with creating training datasets for machine learning applications from experimental high energy and nuclear physics data is separating contributions from different event sources. For example, when considering a binary classification task aiming to separate signal events from backgrounds. In this case it is imperative that the training sample has reliable samples of the different distributions so that the machine learning algorithm can learn the underlying properties of both signal and background. One typical solution is to train the machine learning model on simulated data where the event sources can be trivially labelled. However this relies on excellent agreement between the simulation and real experiment on all training variables which is not always feasible and can lead to sub-optimal classification.\\

\noindent
The \emph{sPlot}~\cite{sPlot} formalism aims to separate the contributions of different event sources to the experimental data. The data is assumed to be characterized by discriminating variables for which the distribution of all sources of events are known, and control variables for which the distributions of some or all sources of events are unknown. These control variables are the features that would be used to train a machine learning model.\\

\noindent
The \emph{sPlot} technique is therefore an effective tool for separating various event sources, facilitating the creation of training samples from actual experimental data without the need for detector simulations. However, values of \emph{sWeights} may be negative~\cite{sPlot}, which leads to complications when training machine learning algorithms. The weighted binary cross-entropy loss function is defined as:

\begin{equation} \label{eq_logloss}
    L(f(x_{i}))= - \sum\limits_{i} w_{i} ( y_{i} \text{log}f(x_{i}) + (1-y_{i}) \text{log}(1-f(x_{i})) )
\end{equation}

\noindent
where $f(x_{i})$ is the output of a learning algorithm for event $i$ which has features $x_{i}$, weight $w_{i}$ and label $y_{i}$. In a binary classification task, $y_{i}$ is equal to either zero or one depending on which class the event belongs to, for example $y_{i}$ can be set to one (zero) for a signal (background) event. As learning algorithms are trained to minimise the loss, loss functions must have a lower bound, otherwise the loss can then be made arbitrarily low. For example, given the loss in Equation~\ref{eq_logloss} above, for a negatively weighted signal event where $y_{i}=1$ ($1-y_{i}=0$), the output of the learning algorithm could be made arbitrarily close to zero, which would make the loss for that event infinitely small. This is problematic, firstly because the signal event is assigned an output close to zero instead of close to one, and secondly because events with negative weights will then dominate the loss function. In general, loss functions that do not have a lower bound will lead to issues during training and poor performance of the learning algorithm.\\

\noindent
There has been some work in resampling negative weights for Monte Carlo event generators where negative weights are redistributed locally in phase space with any potential bias introduced by the resampling becoming arbitrarily small given sufficient statistics~\cite{CellResampler}. Although for Monte Carlo event generators it is always possible to generate larger datasets, experimental data may be statistically limited due to the availability and cost of data taking opportunities. As such, other approaches to resampling negative weights are desirable.\\

\noindent
The use of machine learning to negate the impact of negative \emph{sWeights} has previously been investigated for training learning models in classification tasks~\cite{MLsPlot1,MLsPlot2}. This approach is akin to a regression problem where neural networks or Catboost decision trees~\cite{catboost} were trained to learn the signal and background probabilities produced by the \emph{sPlot} technique using the mean square error loss function:

\begin{equation}
    L(f(x_{i}))=\sum\limits_{i}(f(x_{i}) - w_{i})^{2} \text{  .}\\
\end{equation}

\noindent
$x_{i}$ are the control variables over which the signal and background distributions were separated and the output of the neural network $f(x_{i})$ was constrained between zero and one by using an appropriate activation function in the output node such as a sigmoid function~\cite{MLsPlot1}. As such, the model will be able to predict weights between zero and one but will predict negative weights as zero and weights above one as one. When aiming to separate two classes, each contaminated by background events, the constrained weight $f(x_{i})$ was then used to replace the weight $w_{i}$ in the cross-entropy loss function described in Equation~\ref{eq_logloss}. This approach was shown to be useful in training algorithms to distinguish between the sWeighted signal and background sources~\cite{MLsPlot1,MLsPlot2}. However the statistical properties associated with \emph{sWeights} are then lost due to the loss being constrained to avoid negative weights and weights superior to one. For example, the sum of \emph{sWeights} for a given event source must be equal to the yield for that event source. This requirement is unmet when constraining the weights between zero and one.\\

\noindent
The key technique utilised here is to recast the \emph{sWeights} of a given event source to a probability of the event being from that source via the ratio of the density for that source divided by the sum of the densities of all sources. Binary classification can be used to estimate density ratios~\cite{sugiyama_suzuki_kanamori_2012}. Previous works have investigated the use of probability classifier based density ratio estimation. Refs.~\cite{Martschei_2012,PhysRevD.101.091901} used a neural network model to calculate weights used for reweighting Monte Carlo samples to more closely agree with data. Ref.~\cite{rogozhnikov2016reweighting} used a Boosted Decision Tree model with a bespoke objective function to iteratively reweight histograms. Refs.~\cite{OmniFold} and~\cite{macparticlesPaper} used neural networks and boosted decision trees to model the detection efficiency and acceptance of high energy and nuclear physics experiments from simulations of the experiments.\\

\noindent
In Ref.~\cite{NeuralResampler} a neural network was used to resample positive and negative weights produced by a Monte Carlo event generator using density ratio estimation. The neural network was trained to distinguish between two samples with features $x_{i}$: one with weights given from the MC generator $w_{i}$ and the second with weights equal to 1. When considering the weighted binary cross entropy loss defined in Equation~\ref{eq_logloss}, $y_{i} \text{log}f(x_{i})$ goes to zero when $y_{i}=0$ and $(1-y_{i}) \text{log}(1-f(x_{i}))$ goes to zero when $y_{i}=1$. As such, when the second sample with $y_{i}=0$ is weighted by 1,  the weighted binary cross entropy loss can then be rewritten as:

\begin{equation} \label{eqn:neuralresampler_loss}
    L(f(x_{i}))= - \sum\limits_{i} (w_{i} y_{i} \text{log}f(x_{i}) + (1-y_{i}) \text{log}(1-f(x_{i})) ) \text{  .}\\
\end{equation}

\noindent
This loss function avoids issues due to negative weights encountered in Equation~\ref{eq_logloss} so long as the sum of the weights from the MC event generator is less negative than the number of events produced by the event generator. That is to say that in the case where the sum of weights is negative, its absolute value must be smaller than the number of events produced by the event generator. Ref.~\cite{NeuralResampler} focused on resampling Monte Carlo weights, while pointing out this was applicable for any negative weights application. Here we specifically derive a similar approach to convert \emph{sWeights} to probabilities, while the technique is again more general for the case of negative weight applications.\\

\noindent
This article will demonstrate how \emph{sWeights} can be transformed to event probabilities using density ratio estimation. This article will also demonstrate how decision trees are ideally suited to learn \emph{sWeights}. For certain applications, decision trees are preferable to neural networks as decision trees can have increased computational prediction rates whilst requiring almost no hyperparameter optimisation. Decision trees also typically do not require as large training datasets as neural networks, which can be beneficial when the amount of experimental data is limited. It is therefore beneficial to have flexibility in the choice of learning algorithm. In addition, this article will discuss the biases and correlations that are introduced in the data when using a learning algorithm to convert negative weights to positive definite probabilities, along with demonstrating that these biases are acceptable for the purposes of creating machine learning training data sets from experimental data. \\

\noindent
The rest of the article is organised as follows: Section~\ref{sect_method} will review the \emph{sPlot} formalism before describing how decision trees cope with negative weights. The remainder of Section~\ref{sect_method} will describe how density ratios can be used to convert \emph{sWeights} to probabilities. Section~\ref{sect_casestudies} will then use three case studies, two based on a toy data set and one based on experimental data taken with the CLAS12 experiment~\cite{C12Overview}, to demonstrate the good performance of the method. Section~\ref{sect_ccl} ends the article with brief conclusions and perspectives.\\

\section{Methodology} \label{sect_method}

\subsection{Summary of sPlot}
\noindent
The \emph{sPlot}~\cite{sPlot} technique allows one to disentangle event distributions of different species from a data sample via a discriminatory variable, in which the different species have different distributions of known type. A common use case is to remove background to leave a signal only distribution. This is its purpose in this work where we wish to train classifiers on signal distributions from real experimental data. Essentially, \emph{sPlot} generalises side-band subtraction weights to situations where there is no clear region of isolated background which can be used to subtract from the total event sample. Similar to side-bands it requires that the discriminatory variable and variables of interest are independent of each other. A further generalisation of \emph{sPlot} to treat cases where variable dependence arises is suggested in Custom Orthogonal Weight functions for Event Classification~\cite{DEMBINSKI2022167270}. This work also provided the implementation of \emph{sWeights} used here~\cite{cows}.\\

\noindent
The \emph{sPlot} technique allows one to reconstruct variables' of interest distributions for each event source using the probability density functions (pdf) of each, often established by fitting the expected pdfs on the discriminating variables. In effect, the behavior of the individual sources of events with respect to the variables of interest is inferred from the knowledge available for the discriminating variables by assigning an \emph{sWeight} to each event in the data sample.\\

\noindent
One essential characteristic of \emph{sWeights} is that weights used to remove background species tend to be negative. On one hand this allows the statistical properties of the disentangled data-set to be robust, i.e. uncertainties can be reliably determined from \emph{sWeights} subtracted data provided these weights are propagated to the uncertainties appropriately~\cite{Langenbruch2022}. On the other hand this provides issues for machine learning applications as described in the introduction.\\

\subsubsection{Decision Trees} \label{sect_methodAlgo}
\noindent
For our probabilistic classification task we chose to test a selection of Decision Trees as these naturally accept negative sample weights, such as \emph{sWeights}, when trained with the binary cross entropy loss function of Equation~\ref{eqn:neuralresampler_loss}. Decision trees are composed of nodes which branch out into child nodes. The last node at the end of a branch is known as a leaf and returns a prediction, which in a classification task should be one of the classes in the training data. Decision trees attempt to classify the training data by repeatedly splitting the training data into the left and right child nodes. The splits are performed by applying simple requirements on a random choice of input features such that a specified loss function is minimised. The splitting continues until either all leaves contain one event each or a specified maximum depth is reached. The prediction rate of decision trees can be increased by discretising the input feature space, allowing the decision tree to operate on a bin value rather than specific values of the input features. This type of decision tree is called a histogram decision tree. \\

\noindent
One common issue with decision trees is that they are prone to overfitting the training data. One simple solution to reduce overfitting and generally improve the performance of decision trees is so called boosting. The idea behind boosting is that it is generally easier to train several smaller models than a single large model whilst still avoiding over-fitting. Popular boosting algorithms include adaptive boosting~\cite{ADABoost} or gradient boosting~\cite{GradBoost}. The boosting algorithm employs multiple decision trees trained one after the other, with subsequent decision trees focused on events incorrectly classified by previous decision trees by assigning a suitably defined weight, $w > 1$, to such events. A user defined maximum number of decision trees or a threshold on the training error will stop the boosting process. \\

\noindent
The default node splitting loss in the scikit-learn (version 1.5.0)~\cite{scikit-learn} implementation of gradient boosted decision trees is the weighted binary cross-entropy loss. The loss function can be made to avoid issues in training due to negative \emph{sWeights} by creating a training sample with one class weighted with \emph{sWeights} and the other with weights set to 1 as defined in Equation~\ref{eqn:neuralresampler_loss} and explained in the introduction. The output of a decision tree differs from neural networks in that the output of a leaf depends on the proportion of a class $k$ in a leaf $m$~\cite{treeimpl}. For an unweighted classification task, the output is given by:

\begin{equation}
    p_{mk}=\frac{N_{mk}}{N_{m}}
\end{equation}

\noindent
where $N_{m}$ is the number of events in leaf $m$ and $N_{mk}$ is the number of events in leaf $m$ belonging to class $k$. For a binary classification task with two classes, the ratio between the proportion of both classes $p_{m1}$ and $p_{m2}$ determines the prediction decision. If $\frac{p_{m1}}{p_{m2}} > 1$, then the prediction for events that reach this leaf will be that they belong to class 1.\\

\noindent
For a weighted binary classification task, the sample weight for the first class is $S^{+}_{1} - S^{-}_{1}$, the difference between the sum of the positive weights $S^{+}_{1}$ and the sum of the absolute value of negative weights $S^{-}_{1}$. If the sample weight for the second class is $S_{2}$, the proportion of each class in a node will be: 

\begin{equation} \label{sect_propweights}
    p_{1}=\frac{S^{+}_{1} - S^{-}_{1}}{N} \text{   and   } p_{2}=\frac{S_{2}}{N}
\end{equation}

\noindent
where the subscript $m$ was dropped for simplicity. The ratio $p_{1}/p_{2}$ from Equation~\ref{sect_propweights} is unchanged when removing negatively weighted events from the first class and adding positively weighted samples to the second class with weights $S^{+}_{2}$ such that:

\begin{align}
    \frac{S^{+}_{1} - S^{-}_{1}}{S_{2}} &= \frac{S^{+}_{1}}{S_{2}+S^{+}_{2}} \notag\\
   \Rightarrow S^{+}_{2} &= \frac{S_{2}S^{-}_{1}}{S^{+}_{1} - S^{-}_{1}} \text{  .} \label{eq_posweight}\\\notag
\end{align}

\noindent
In short, negative weights can be used to train scikit-learn decision trees with the right loss function as negative weights in a given sample are, in effect, akin to adding positively weighted events to the other sample.\\


\subsection{Learning Weights using Density Ratios} \label{sect_methodDR}

\noindent
The aim of the \emph{sPlot} formalism is to separate the true distribution of one or more control variables $x_{i}$ for events of different sources using the knowledge available from discriminating variables on the distribution of these sources. The source are classed in species (e.g. signal and background). The \emph{sPlot} technique provides a consistent representation of how all events from the species are distributed in $x_{i}$~\cite{sPlot}. Summing the \emph{sWeights} for a given species then recovers the yield of that species obtained by a fit to the discriminating variable. For example, by summing the signal weights one recovers the signal yield. Summing the weights for all species allows to recover the entire data distribution composed of the different species.\\

\noindent
The \emph{sWeights} for a given species can then be taken as the ratio of the probability density for that species over the sum of probability densities of all species in the data $D_{all}(x_{i})$. For a distribution separated into signal and background species with probability densities $D_{S}(x_{i})$ and $D_{B}(x_{i})$ respectively, the density ratio weights \emph{drWeights} $W_{dr}(x_{i})$ distribution over control variables $x_{i}$ is written as the density ratio:

\begin{equation}
   W_{dr}(x_{i})=\frac{D_{S}(x_{i})}{D_{S}(x_{i})+D_{B}(x_{i})} = \frac{D_{S}(x_{i})}{D_{all}(x_{i})} \text{  .}\\
\end{equation}

\noindent
The knowledge of this ratio is sufficient to model the signal \emph{sWeights} distribution. Note that this ratio would also be preserved in the presence of more than one background species, as the distribution of all events will simply be expanded to a sum of the signal distribution and all background species' contribution: $D_{all}(x_{i})=D_{S}(x_{i})+D_{B1}(x_{i})+D_{B2}(x_{i})...$ A convenient technique for density ratio estimation is to treat it as a binary classification problem. Similarly to the neural resampler of Ref.~\cite{NeuralResampler}, a machine learning model can be trained on data separated into two classes: the first with density distribution $D_{S}(x_{i})$ consisting of all events in the data weighted with the signal \emph{sWeights}. The second with density distribution $D_{all}(x_{i})$ consisting of all events uniformly weighted with weights set to one. $D_{S}(x_{i})$ is labelled class 1 and $D_{all}(x_{i})$ is labelled class 0. The output of the classifier $f(x_{i})$ for class 1 is then:

\begin{align}
    \notag f(x_{i}) &= \frac{D_{S}(x_{i})}{D_{S}(x_{i}) + D_{all}(x_{i})} =\frac{D_{S}(x_{i})}{D_{S}(x_{i}) + D_{S}(x_{i})+D_{B}(x_{i})}\\ \notag
    \Rightarrow \frac{1}{f(x_{i})} &= \frac{D_{S}(x_{i})}{D_{S}(x_{i})} + \frac{D_{S}(x_{i})}{D_{S}(x_{i})} + \frac{D_{B}(x_{i})}{D_{S}(x_{i})}\\ \notag
    \Rightarrow \frac{1}{f(x_{i})} - 1 &= \frac{1-f(x_{i})}{f(x_{i})} = \frac{D_{S}(x_{i})+D_{B}(x_{i})}{D_{S}(x_{i})}\\ 
    \Rightarrow W_{dr}(x_{i}) &= \frac{f(x_{i})}{1-f(x_{i})} \text{  .}\\\notag
\end{align}

\noindent
Overall, the two key aspects of converting \emph{sWeights} to probabilities using density ratio estimation are first that creating the training sample with one class weighted by the \emph{sWeights} and the other with weights set to one allows to use the binary cross-entropy loss function without issues in training due to negative \emph{sWeights} as it preserves a lower bound in the loss function. Second, creating the training sample in such a way allows a binary classification model to learn the ratio of the signal probability density divided by the sum of the probability densities of all species, which is equal to the signal \emph{sWeights} distribution. Binary classification for density ratio estimation is therefore perfectly suited to convert \emph{sWeights} into probabilities. Note that the learning algorithm should only be trained with control variables as inputs and not discriminatory variables. \\

\noindent
Section~\ref{sect_methodAlgo} demonstrated how the scitkit-learn implementation of boosted decision trees allows training with negative weights. In the schema proposed above, negative weights in class 1, with density distribution $D_{S}$, would have the same effect as adding positive weights to class 0, with density distribution $D_{all}$. The density ratio which describes \emph{drWeights} $W_{dr}$ is preserved by boosted decision trees, as removing negative weights from $D_{S}$ and adding the corresponding positive weights to $D_{all}$ must preserve the overall ratio. \\

\noindent
A final consideration is to preserve the uncertainty associated with using the \emph{sWeights}. This uncertainty generally requires taking the sum of the squared \emph{sWeights}. However, Ref.~\cite{NeuralResampler} showed that calculating the uncertainty as the sum of the resampled weights squared gives incorrect uncertainties. There are two possible solutions; the first is to to estimate the uncertainties by the quadratic sum of the \emph{sWeights}. Second, Ref.~\cite{NeuralResampler} also showed that the uncertainty itself can be converted using density ratio estimation, which preserves the uncertainty. For the remainder of this article, we choose the first option, to carry over the uncertainty from \emph{sWeights}, but it is important to note that the uncertainty on \emph{sWeights} can also be learned.\\

\section{Case Studies} \label{sect_casestudies}

\subsection{Toy Example} \label{sect_toy}
\noindent
This section will present a toy example to illustrate the performance of the density ratio estimation of \emph{sWeights}. In the toy example, a simple event generator produces three dimensional events where the first variable is akin to a mass such as the invariant mass of a given reaction, the second variable is akin to an azimuthal ($\phi$) angular distribution and the third variable is $z=\cos{\theta}$. Signal events were generated with a Gaussian distribution in mass and a $\cos{2\phi}$ of amplitude 0.8. Background events were generated with a linear mass distribution and a $\cos{2\phi}$ of amplitude -0.2. The ratio of signal to background was varied in different tests. The mass variable was used as the discriminatory variable, allowing us to separate the signal and background distributions via the weights. The purpose of this toy example was then to measure the amplitude of the signal in $\phi$ by separating the signal and background distributions in the control variables $\phi$ and $z$.\\

\begin{figure}[ht!]
    \centering   
    \includegraphics[width=0.49\textwidth]{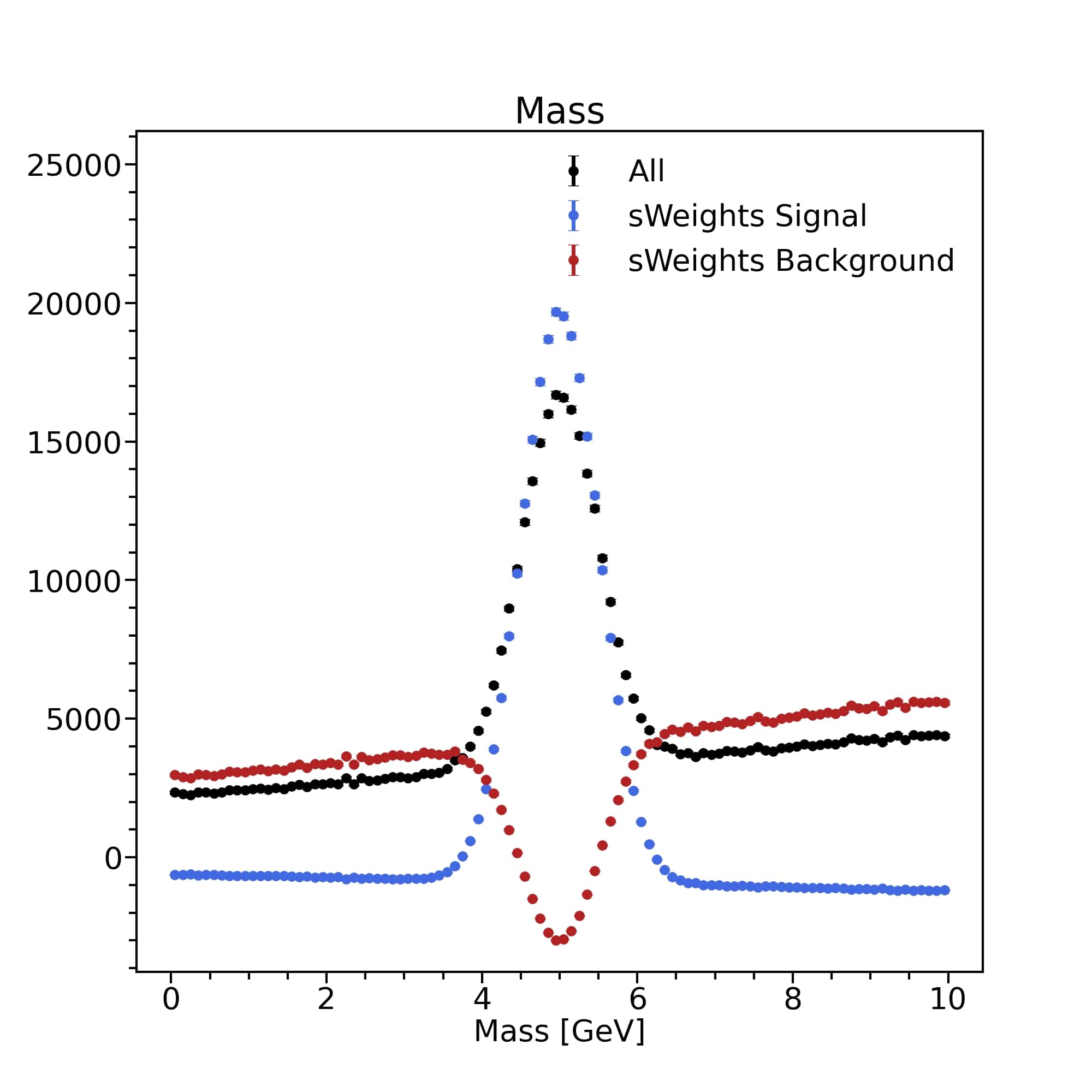}
    \includegraphics[width=0.49\textwidth]{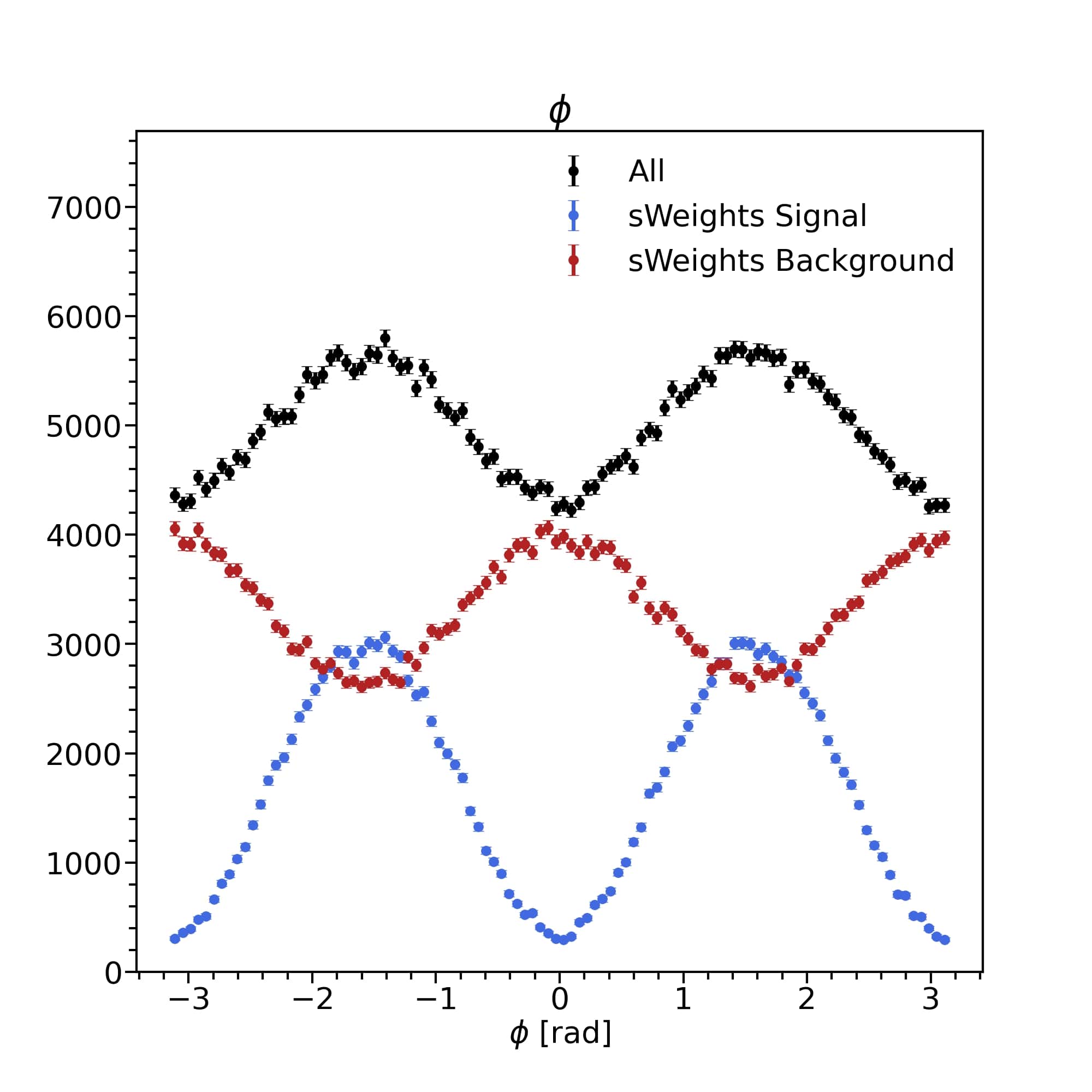}
    \includegraphics[width=0.49\textwidth]{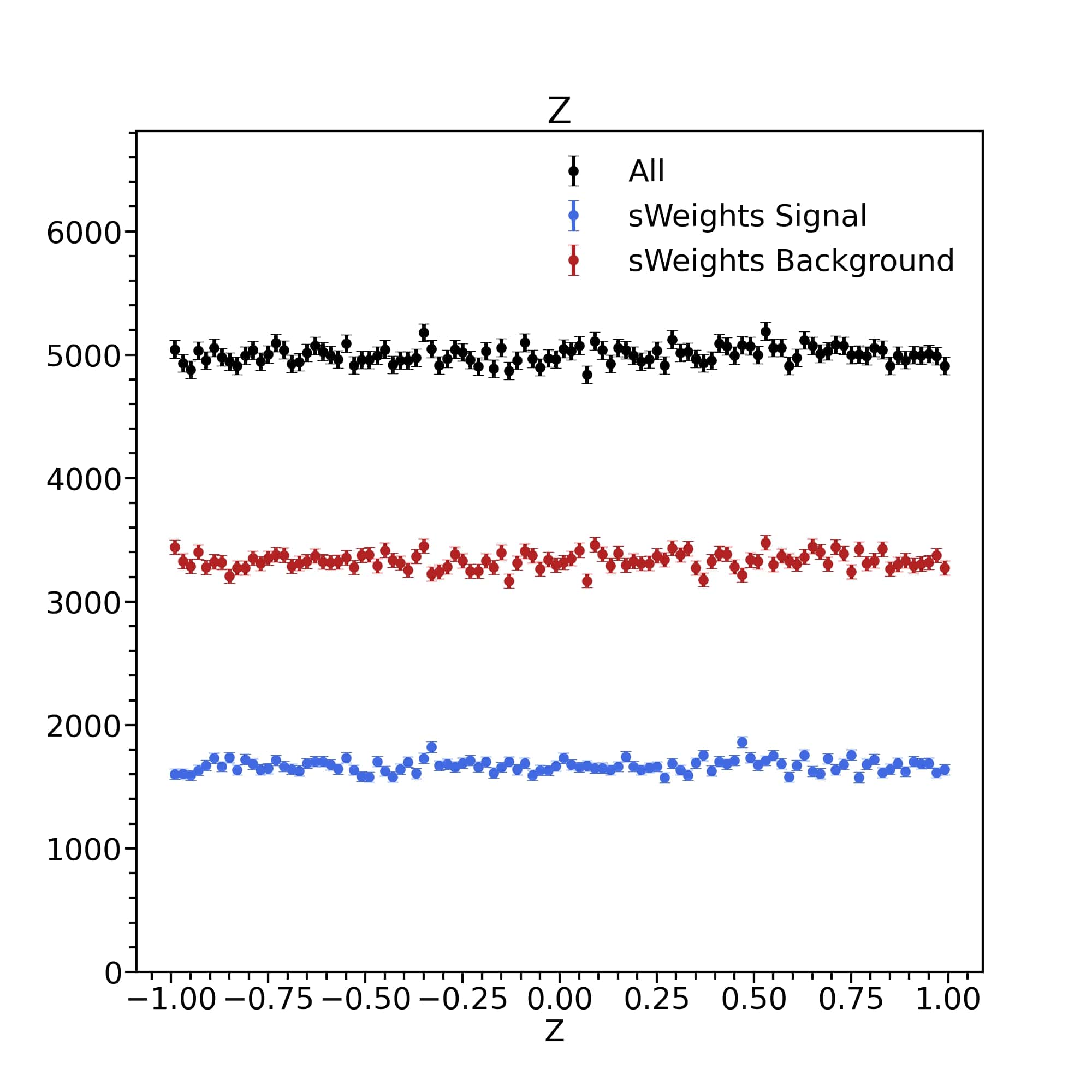}
    \caption[Comparing the toy total, signal and background distributions.]{The total (black) distribution in mass (left), $\phi$ (middle) and Z (right) compared to the same distribution with signal or background \emph{sWeights} applied.}
    \label{fig:toy_sWeights}
\end{figure}

\noindent
The generated mass distribution was fitted using the sum of the signal and background pdfs used to generate the events with the mean, width, and polynomial coefficients allowed to vary in the fit. The fit was then used to calculate the signal and background \emph{sWeights} using the implementation from Ref.\cite{DEMBINSKI2022167270,cows}. Figure~\ref{fig:toy_sWeights} shows a comparison of the mass, $\phi$ and Z distributions for all events and these same events with applied signal and background \emph{sWeights}, with a signal-to-background ratio of 1:2. As can be seen, the sum of the signal and background weights allows to reproduce the total distribution. The discriminatory variable, mass, contains negative bins for the signal and background distribution, where these events are effectively subtracted to give the disentangled control distributions.\\

\noindent
The methodology described in Section~\ref{sect_method} was then applied to convert the \emph{sWeights} using density ratio estimation. The scikit-learn library (version 1.5.0)~\cite{scikit-learn} was used to test both a gradient boosted decision tree (GBDT) and a histogram gradient boosted decision tree (HistGBDT). Both the GBDT and HistGBDT were given a maximum depth of 10 and otherwise default parameters. We found that performance improved with depth up to 10, but not significantly after that. The neural resampling method described in Ref.~\cite{NeuralResampler} was also applied to the toy example. Ref.~\cite{NeuralResampler} used particle flow networks (PFN)~\cite{PFN} based on the deep sets architecture~\cite{deepsets}. PFNs are general models designed for learning from collider events as unordered, variable-length sets of particles, rendering them unsuited to this application with only two variables. Instead, we used neural networks (NN) with 7 hidden layers and 1024, 512, 256, 128, 64, 32, 16 nodes, respectively, implemented using tensorflow (version 2.16.1)~\cite{tensorflow}. The hidden layers all had RELU activation functions, the output layer had a sigmoid activation function, and the network was trained with an ADAM optimizer~\cite{adam} for 50 epochs. Ref.~\cite{macparticlesPaper} found that a second iteration of the density ratio estimation, essentially using the density ratio estimation for a reweighting step, improved the performance by fine-tuning the model. This was also tested here. \\

\noindent
The idea behind using a second reweighting step is that after the initial drWeights application there may still be some residual differences between the drWeighted and sWeighted distributions. A second classification may then be applied between the drWeighted and sWeighted distributions. If the two distributions agree, then the weights from this second classification should all be close to 1, but if there are regions where the agreement is not so close, the second weights will correct the first. This second iteration weighted the class labeled 0 with the weights produced by the first density ratio estimation, and the predicted weights were then taken as the product of the weights obtained by both individual models.\\

\noindent
Training and prediction times were estimated using 5 cores of an AMD EPYC 9554 64-Core Processor at 3.1 GHz. $3.5\times10^{5}$ events were generated, leading to a training sample with $7\times 10^{5}$ events since the same events are found in both classes 1 and 0, albeit with and without \emph{sWeights} weighting. The GBDTs train at a rate of 4kHz and had a prediction rate of roughly 1.6 MHz. The HistGBDTs had a training rate of roughly 1 MHz, with a prediction rate of roughly 20 MHz. The neural network described above had a training rate of 2 kHz with a prediction rate of 0.5 MHz. The impact of the number of training events is discussed later in the text. Unless otherwise mentioned, tests were made with $10^5$ generated events, leading to training samples with $2\times 10^{5}$ events since the same events are found in both classes 1 and 0 with or without \emph{sWeight} weighting.\\

\noindent
The $cos(2\phi)$ amplitude was then extracted via the weighted datasets. To do this a 1D histogram was generated in $\phi$ with 100 bins, each bin filled with the relevant signal \emph{sWeights} or converted weights called \emph{drWeights}. While the \emph{drWeights} reproduced the correct distribution, the uncertainties were calculated from the sum of the original \emph{sWeights} in each bin to allow correct propagation of errors to the fit result as discussed at the end of Section~\ref{sect_methodDR}. Fits were performed by minimizing the $\chi^2$ between the model and the binned histogram using the \emph{iminuit} package~\cite{iminuit}.\\

\noindent
If the \emph{sWeights} and \emph{drWeights} accurately separate the signal and background distributions in $\phi$ then the amplitude obtained by the fit should be consistent with the generated value of 0.8. Several different converter models were tested along with a density ratio product estimation where the first model was fine tuned by the second model. The total number of generated events was $10^5$ with a signal to background ratio of either 1:2 or 1:9. All events were used in training. The entire chain of generating data, fitting and calculating \emph{sWeights}, training the density ratio model and measuring the $\cos(2\phi)$ amplitude was repeated 50 times to allow us to determine the robustness of the conversion procedure. We report the mean amplitude and uncertainty $\Bar{\sigma}_{fit}$ from the 50 \emph{iminuit} fits and the standard deviation, $\hat{\sigma}_{rms}$, of the amplitude over the 50 data sets. The expectation is that the mean should be consistent with the nominal value of 0.8, while $\Bar{\sigma}_{fit}$ and $\hat{\sigma}_{rms}$ should be numerically similar, i.e. the fluctuation of results is consistent with the calculated uncertainty.\\

\noindent
The mean amplitude, $\Bar{\sigma}_{fit}$, and $\hat{\sigma}_{rms}$ of the amplitude over 50 iterations when varying several parameters of the \emph{drWeights} are contained in~\ref{app_tests} for conciseness. Table~\ref{tab:toy_depth} shows the performance of a single GBDT converter when varying the maximum depth of the GBDT, demonstrating that  deeper GBDTs produce better results as they fully capture correlations in the input variable space. A maximum depth of 10 is chosen as further increasing the depth only leads to longer training and prediction times without increasing the performance of the method. Table~\ref{tab:toy_models} shows the results from different classifiers such as a HistGBDT, a GBDT and a neural network. The GBDT produces the most reliable results, and in particular adding a reweighting step or density ratio product produces the best performance. Most converters are able to reproduce the nominal amplitude of 0.8 but $\hat{\sigma}_{rms}$ varies significantly from $\Bar{\sigma}_{fit}$ for some converter models. However, the density ratio product using two GBDTs is able to reproduce the nominal amplitude of 0.8 for both signal to background ratios of 1:2 and 1:9 and obtains consistent $\hat{\sigma}_{rms}$ and $\Bar{\sigma}_{fit}$, proving to be the most robust converter. Table~\ref{tab:toy_nEvs} shows the performance of the GBDT density ratio product when varying the number of generated events. In all cases the amplitude measured with the \emph{drWeights} is consistent with the generated amplitude of 0.8. The $\Bar{\sigma}_{fit}$ and $\hat{\sigma}_{rms}$ on the amplitudes are generally consistent. The good performance of the \emph{drWeights} even with large backgrounds and small event samples is a key consideration, given that experimental datasets may be limited and may have irreducible backgrounds. Although it is worth emphasizing that \emph{drWeights} should only be used when positive definite probabilities are required as the \emph{sWeights} proved to be generally more reliable, overall the \emph{drWeights} are found to be a robust conversion of \emph{sWeights} to probabilities.\\

\begin{figure}[ht!]
    \centering   
    \includegraphics[width=0.49\textwidth]{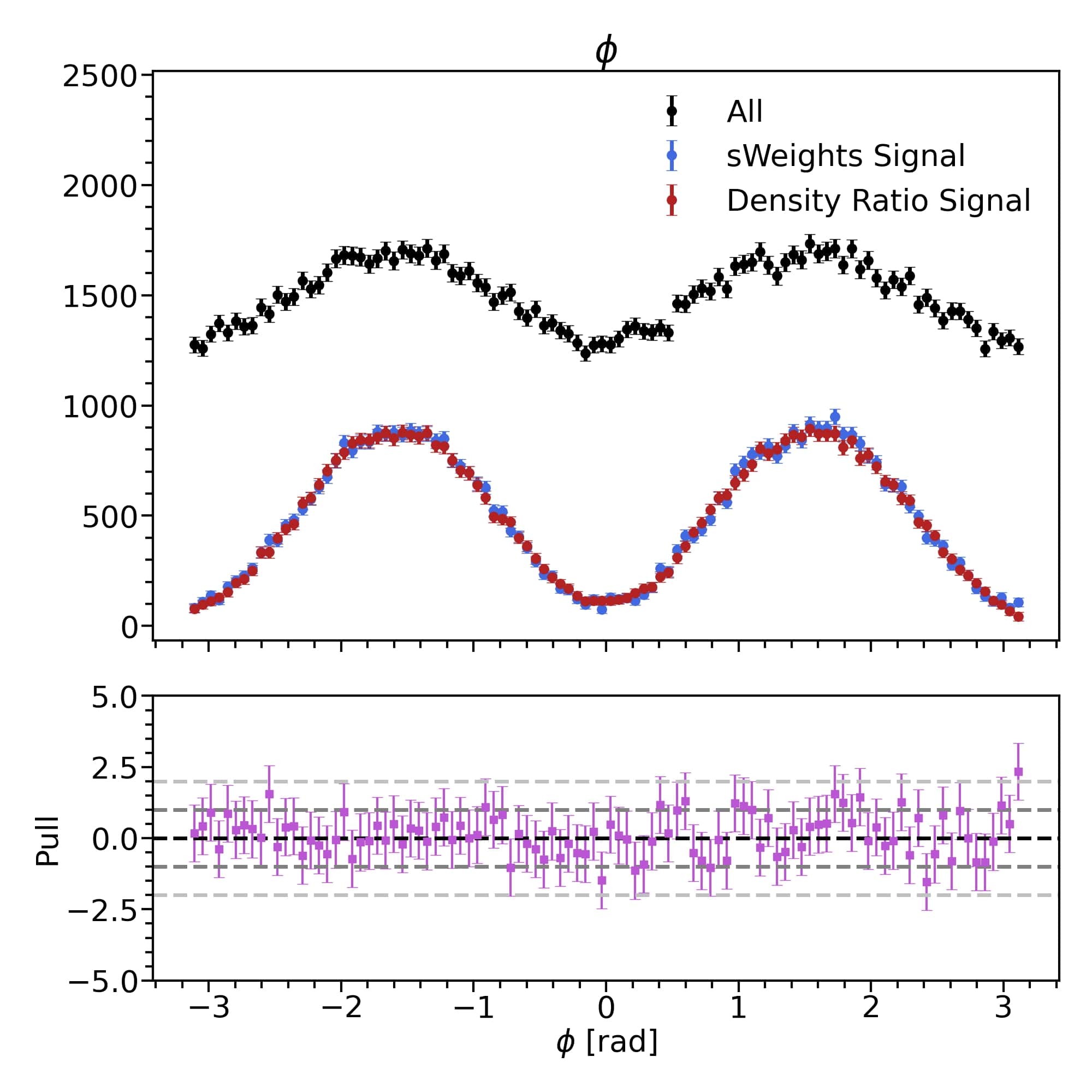}
    \includegraphics[width=0.49\textwidth]{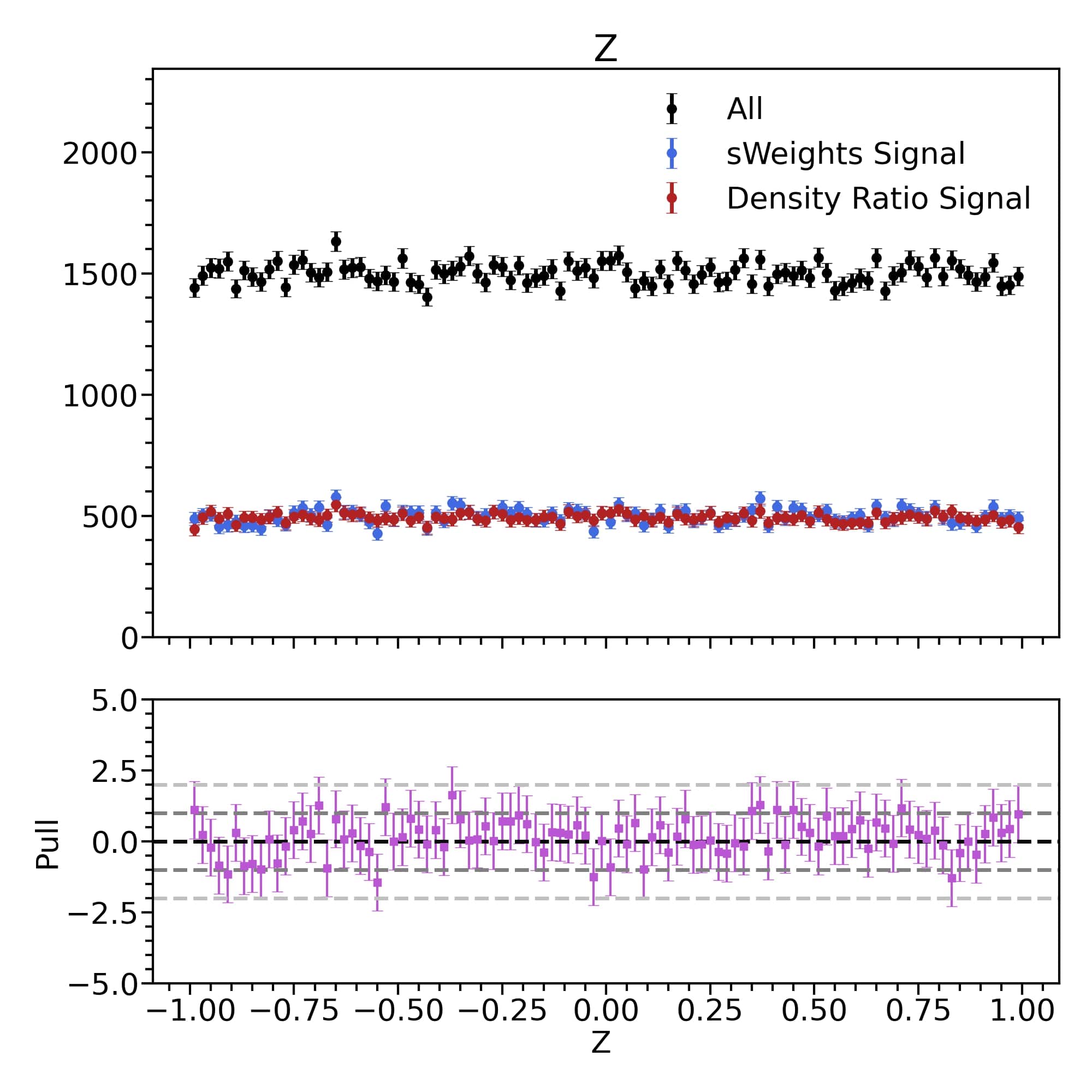}\\
    \includegraphics[width=0.49\textwidth]{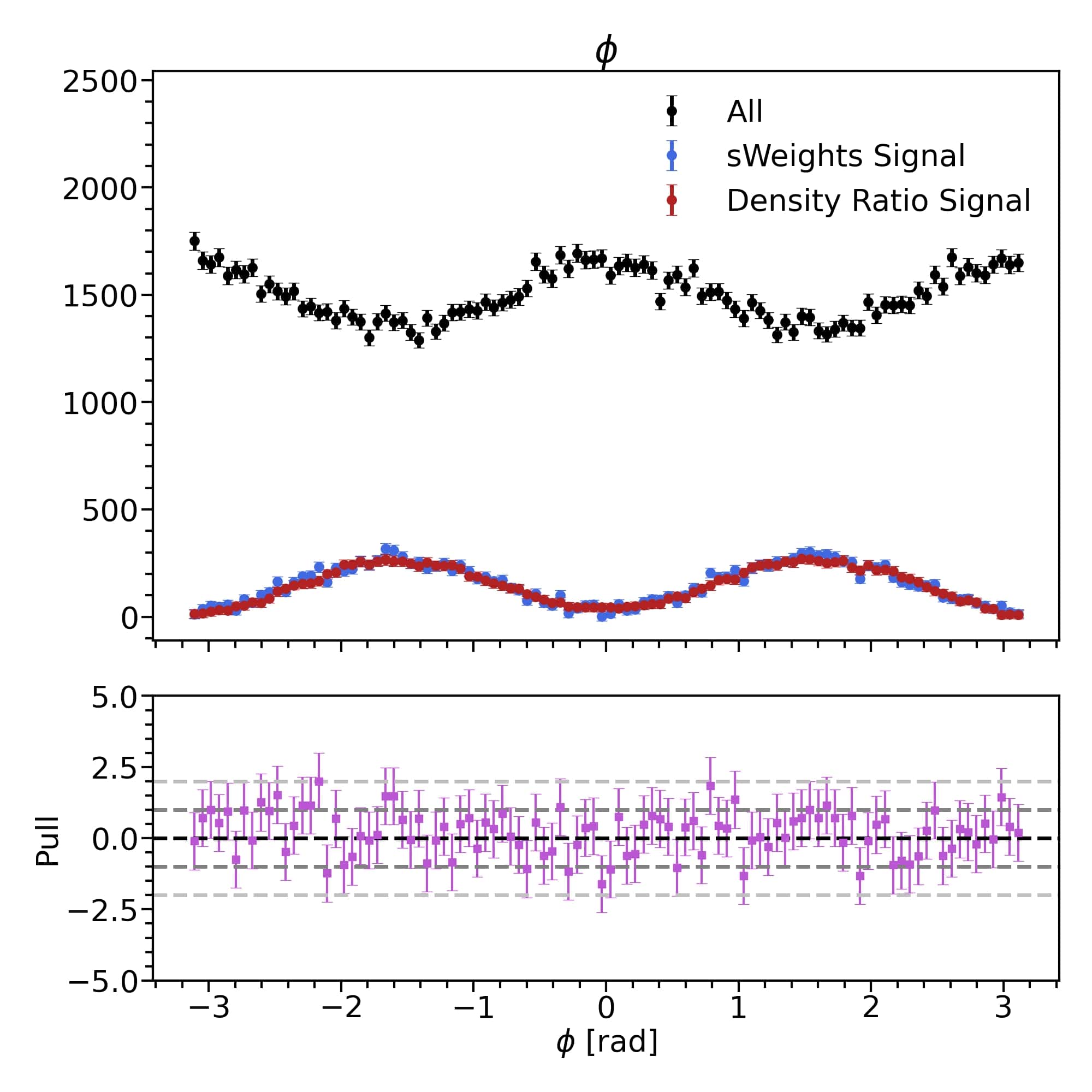}
    \includegraphics[width=0.49\textwidth]{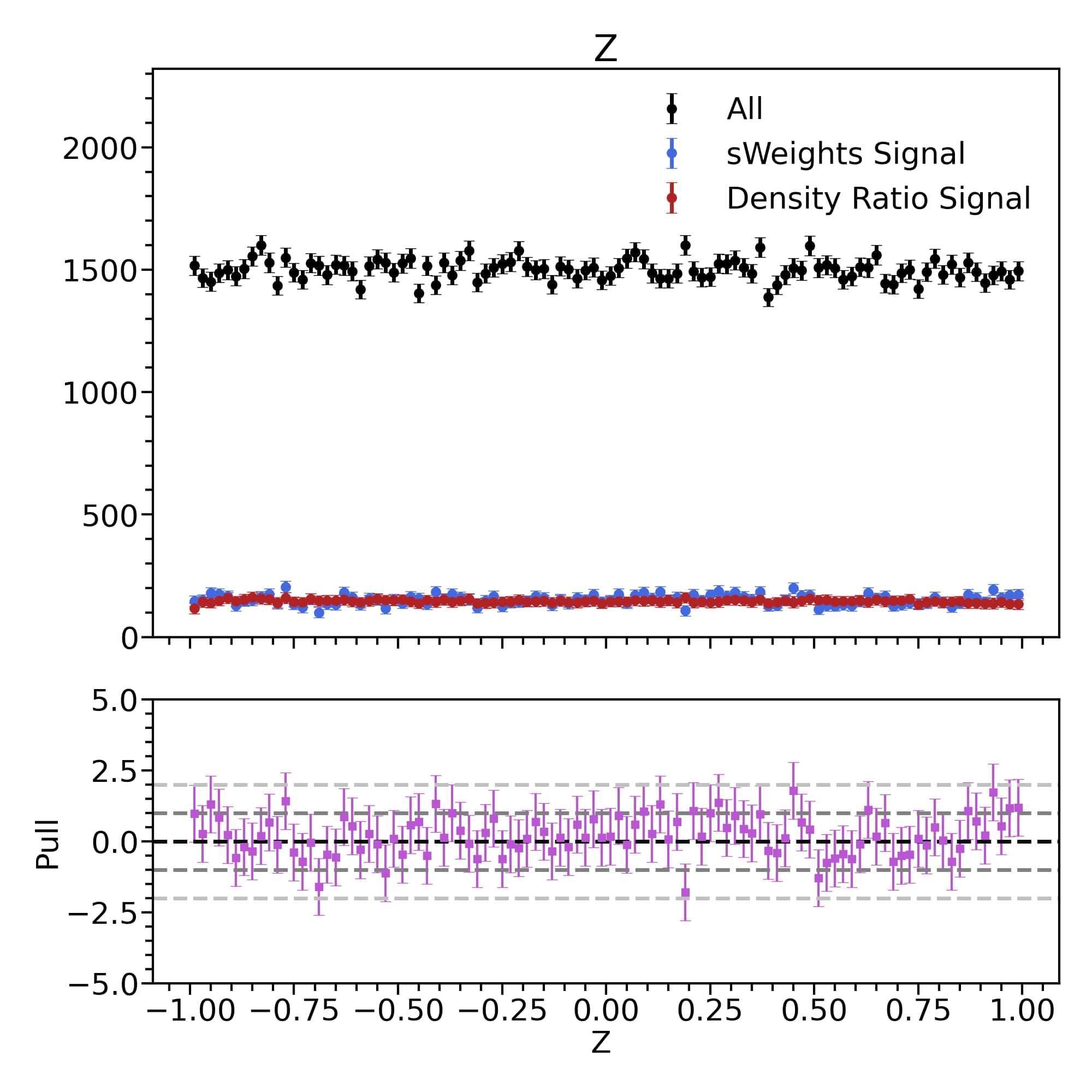}
    \caption[Comparing the \emph{sWeights} to DR Weights in toy example.]{The total (black) distribution in $\phi$ (left) and Z (right) compared to the same distribution with signal \emph{sWeights} applied (blue) or the \emph{drWeights} applied (red), for $5\times10^5$ generated events with 0.35M used in training and 0.15M plotted here. The top row was generated with a signal to background ratio of 1:2, whereas the bottom row has a signal to background ratio of 1:9. The pull, $g_{hist}$, between the \emph{sWeights} and \emph{drWeights} distributions is also shown below all distributions. The \emph{drWeights} are estimated using the GBDT and GBDT density ratio product.}
    \label{fig:toy_results}
\end{figure}

\noindent
Figure~\ref{fig:toy_results} shows a comparison of the signal distributions separated from background in $\phi$ and $z$ using the signal \emph{sWeights} and \emph{drWeights} for the best algorithm, namely the density ratio product using the GBDT followed by another GBDT. Figure~\ref{fig:toy_results} also compares the case where the events were generated with a signal to background ratio of 1:2 or 1:9. A total of 0.5M events were generated events where 0.35M were used in training and 0.15M were used for testing and are plotted in Figure~\ref{fig:toy_results}. The uncertainties are calculated from the original \emph{sWeights}. Figure~\ref{fig:toy_results} also shows the bin-by-bin pull distributions defined as the difference between the bin values for the two distributions divided by the sum squared of their uncertainties. As the uncertainties are equal for both histograms, given that they come from the sum of the \emph{sWeights} squared, the pull calculation simplifies to:
\begin{equation}
    g_{hist} = \frac{S_i - DR_i}{\sqrt{2}\sigma_S }.
\end{equation}
In general, the estimation of the density ratio is capable of accurately reproducing the weighted distributions of the $\phi$ and Z signals, as shown by the pulls. This also holds for a dominant background, as shown in the bottom row of Figure~\ref{fig:toy_results}. For both signal-to-background ratios of 1:2 or 1:9, the $\cos(2\phi)$ amplitudes measured with the signal \emph{drWeights} were consistent within uncertainties with the generated value of 0.8. We note that the mean of the $g_{hist}$ distributions is 0.25 showing a small bias consistent with a 1$\%$ reduction in the \emph{drWeighted} distribution, while the mean widths are less than 1 at around 0.45, showing that the two distributions are not statistically independent even if the \emph{drWeights} do not exactly reproduce the \emph{sWeights}. \\

\noindent
In order to study the statistical properties of the \emph{drWeighted} distributions in more detail, we ran a longer series of 2000 toy test iterations, each with $10^{5}$ generated events and a signal to background ratio of 1:9. The results of this are shown in Table \ref{tab:high_stats}. We see that the \emph{sWeights} perform as expected with the fits producing an unbiased mean with an average fit uncertainty, $\Bar{\sigma}_{fit}$, equal to the standard deviation, $\hat{\sigma}_{rms}$, of the $10^{5}$ results. On the other hand, we see that there is a small bias in the \emph{drWeighted} distribution, with its mean amplitude reduced by $1\%$ and $\hat{\sigma}_{rms}$ increasing moderately by $15\%$. The bias is around one-quarter of a standard deviation.
Further we quote the mean reduced $\chi^{2}$ of the 2000 fits; again we see that the \emph{sWeighted} distributions have values close to 1 as expected, while the \emph{drWeighted} distributions are significantly lower at 0.66. This is a result of bin-by-bin smoothing resulting from the converter, which must effectively average over neighboring events to assess the density ratio. This can be seen in Figure \ref{fig:toy_results} where the \emph{drWeights} distributions have less statistical fluctuations. We also show the average of the pull means and widths, where the individual pull means and widths were calculated from the weighted histogram content ($H_i$) and fit result ($f_i$) :

\begin{equation}
    g_{fit} = \frac{H_i - f(\phi_i)}{\sigma_i }.
\end{equation}

\noindent
The mean and standard deviation of $g_{fit}$ are then averaged over all 2000 fits. The \emph{sWeights} distributions follow a normal distribution, while the \emph{drWeights} distributions have a significantly lower width of 0.81 (compared to unity), again a result of the smoothing introduced by the density ratio converter. \\

\begin{table}[ht!]
\centering
\begin{tabular}{ |c||c|c|c|c|c|c|c|} 
\hline
Weights & & &  &  & & &\\
 & Amp. Mean & $\hat{\sigma}_{rms}$ & $\Bar{\sigma}_{fit}$ & $\frac{ \hat{\sigma}_{rms}}{\Bar{\sigma}_{fit}}$ & $\chi^{2}$ & $\Bar{g_{fit}}$ & $\Bar{\sigma}_{g_{fit}}$ \\
\hline
\emph{sWeights}& & &  & & & &\\
 & 0.801 & 0.027 & 0.027 & 1.01  & 0.94 & 0.02 & 0.97 \\
\hline
\emph{drWeights} & & &  &  & & &\\
 & 0.792 & 0.031 & 0.027 &  1.15 &0.66& 0.01 & 0.81 \\
\hline
\end{tabular}\\
 \caption[Comparison of the $\phi$ amplitude extraction using \emph{sWeights} and \emph{drWeights}.]{Comparison of the $\phi$ amplitude measured with the signal distribution generated with a $\phi$ amplitude of 0.8 for $10^5$ events with a signal to background ratio of 1:9. The data generation and training were repeated 2000 times. The mean and standard deviation ($\hat{\sigma}_{rms}$) of the measured amplitudes are reported, along with the mean fit uncertainty. We also show the mean $\chi^{2}$ of the 2000 fits and the average of the pull ($g_{fit}$) means and standard deviations of the histogram to the fit result, with each mean and standard deviation constructed from the 100 bins of the test histogram.}
    \label{tab:high_stats}
\end{table}

\begin{figure}[ht!]
    \centering   
    \includegraphics[width=0.99\textwidth]{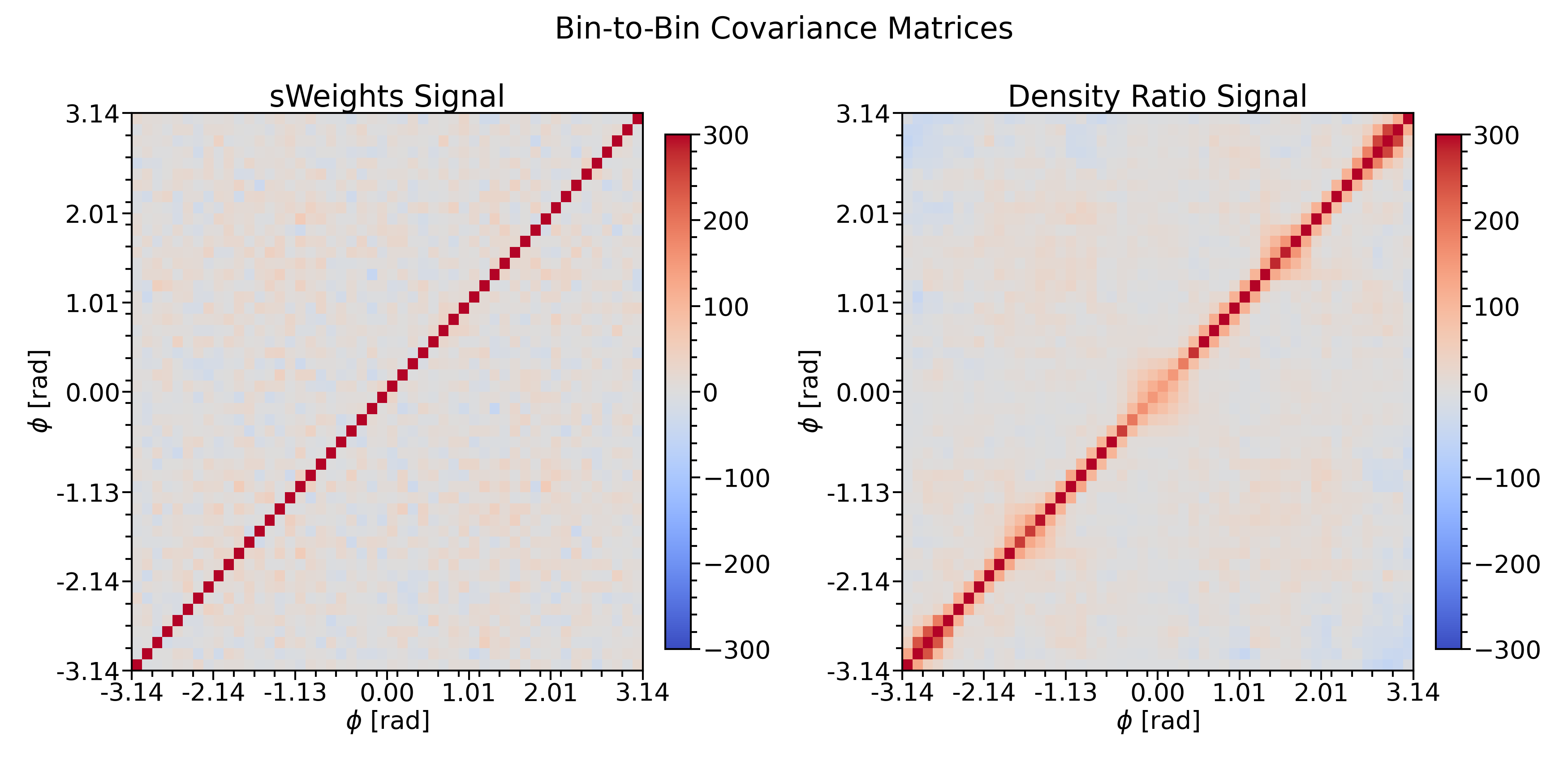}
    \caption[The \emph{sWeight} and \emph{drWeights} bin-to-bin covariance matrices.]{The \emph{sWeight} and \emph{drWeights} bin-to-bin covariance matrices in $\phi$ estimated over 2000 test iterations. The colour axis is capped between -300 and 300 for visualisation purposes but the actual maximal covariance was of order 800.}
    \label{fig:toy_cov}
\end{figure}

\noindent
To further investigate the effect of converter smoothing, we construct a covariance matrix for the two types of weighted distribution. This is done by saving the individual histogram bin contents for each test, such that we have a dataset of 100 by 2000 entries. The \emph{numpy.cov} routine~\cite{NumpyCov} was then used to produce a 100x100 matrix from these data. The covariance matrices for \emph{sWeights} and \emph{drWeights} are shown in Figure~\ref{fig:toy_cov}. Note that, for visualisation purposes, the colour axis was capped to $\pm 300$ whereas the actual maximal covariance was of order 800. The \emph{sWeights} matrix is shown to be diagonal as expected for histograms with uncorrelated bins. Bins around the peaks and troughs of the distribution in the \emph{drWeights} matrix have a significant correlation with their nearest neighbors which also extends to more distant neighbors. This smoothing of the peaks and troughs is then what leads to the small bias in the fit amplitude. These correlations will also result in $\hat{\sigma}_{rms}$ being larger than $\Bar{\sigma}_{fit}$, as the full covariance matrix is not used in the $\chi^{2}$ calculation. In principle, one could correct for that using the full covariance matrix, although that would need to be calculated numerically similar to what is done here.\\


\noindent
We then used the bootstrap technique in order to reliably extract the uncertainty in the \emph{drWeighted} fits for a single data set: for the chosen data set of $10^5$ events, the same number of events were randomly sampled from that original set. The sampling is done with replacement, meaning that the same event may appear multiple times or never. This random sampling was performed 2000 times each on two separate data sets and the results are shown in Table~\ref{tab:bootstrap}. The expectation is that \emph{sWeights} and \emph{drWeights} reproduce the $\phi$ amplitude in the chosen data set in a given bootstrap sample, following a statistical distribution of 0.8 $\pm \hat{\sigma}_{rms}$, where $\hat{\sigma}_{rms}$ was estimated with independent toy samples and that $\hat{\sigma}_{boot}=\hat{\sigma}_{rms}$. We observe that the mean amplitude is within around one standard deviation of the true value. As before, the $\hat{\sigma}_{boot}$ of the \emph{drWeights} increases relative to the $\Bar{\sigma}_{fit}$. The \emph{sWeights} mean fit reduced $\chi^2$ of the 2000 samples is increased beyond unity as one would expect, due to the bootstrap sampling in each set of events. This variation is also reflected in the pull widths, $\Bar{\sigma}_{g_{fit}}$ increasing beyond 1. We also see that the impact of the bin-by-bin smoothing introduced by \emph{drWeights} is similar to that on independent toy samples. The mean fit reduced $\chi^2$ and the pull widths are closer to 1 for \emph{drWeights} as the increases seen in \emph{sWeights} are canceled to some degree by the smoothing in the density ratio. Overall we demonstrate that \emph{drWeights} are well suited to real world applications where precise statistical evaluations are not necessary.\\

\begin{table}[ht!]
\centering
\begin{tabular}{ |c||c|c|c|c|c|c|c|} 
\hline
Weights & & &  &  & & &\\
 & Amp. Mean & $\hat{\sigma}_{boot}$ & $\Bar{\sigma}_{fit}$ & $\frac{ \hat{\sigma}_{rms}}{\Bar{\sigma}_{fit}}$ & $\chi^{2}$ & $\Bar{g_{fit}}$ & $\Bar{\sigma}_{g_{fit}}$ \\
\hline
\emph{sWeights}& & &  & & & &\\
 & 0.798 & 0.0270 & 0.0269 & 1.00  & 1.88 & 0.03 & 1.38 \\
 & 0.837 & 0.0282 & 0.0278 & 1.01 & 1.78 & 0.04 & 1.34 \\
\hline
\emph{drWeights} & & &  &  & & &\\
 & 0.784 & 0.0322 & 0.0269 &  1.19 & 1.12 & 0.02 & 1.06 \\
 & 0.806 & 0.0320 & 0.0292 & 1.10 & 1.03 & 0.02 & 1.02 \\
\hline
\end{tabular}\\
 \caption[Comparison of the $\phi$ amplitude extraction using \emph{sWeights} and \emph{drWeights} on bootstrap samples drawn from a single data set.]{Comparison of the $\phi$ amplitude measured on 2000 bootstrap samples drawn from data sets with the signal distributions generated with a $\phi$ amplitude of 0.8 for $10^5$ events and signal to background ratios of 1:9. Note that the bootstrapping was done on two different data sets corresponding to the two rows for both \emph{sWeights} and \emph{drWeights}. The mean and standard deviation ($\hat{\sigma}_{boot}$) of the measured amplitudes are reported, along with the mean fit uncertainties. We also show the mean $\chi^{2}$ of the 2000 fits and the average of the pull ($g_{fit}$) means and standard deviations, with each mean and standard deviation constructed from the 100 bins of the test histogram.}
    \label{tab:bootstrap}
\end{table}





\noindent
The toy example presented in this section is contained in the Github repository found at~\cite{DR4sWeight}. A \emph{generator} class produces the total distribution composed of the signal and background distributions, calculates the signal and background \emph{sWeights} and fits the $\cos(2\phi)$ amplitude. A \emph{plotter} class allows to produce the plots shown in this section. A \emph{performance} class allows to run the tests described here. A \emph{trainer} class allows to train the models as described in this article. A training script then allows to train the decision tree based models and the neural resampler. The toy example can be used as example of how to train and deploy the methods described in Section~\ref{sect_method} and in Ref.~\cite{NeuralResampler}.\\

\subsection{Higher Frequency Distributions}
\label{sect_casestudies_highfreq}
\noindent
In Section \ref{sect_casestudies} we showed that for a relatively common and simple problem in physics data analysis the \emph{drWeights} were able to reproduce signal distributions with reasonably well behaved statistical properties. However, given that the underlying GBDT algorithm effectively relies on averaging over local densities, some degree of smoothing, or correlating of events, is to be expected, as was demonstrated at the end of Section \ref{sect_casestudies}. To investigate the limits of the \emph{drWeights} technique, we also performed a toy analysis where the structures in the distributions are sharper and so smoothing effects are likely to be more noticeable. This was done by gradually increasing the frequency of the cosine distribution and performing the same toy analysis.\\

\begin{figure}[ht!]
    \centering   
    \includegraphics[width=0.49\textwidth]{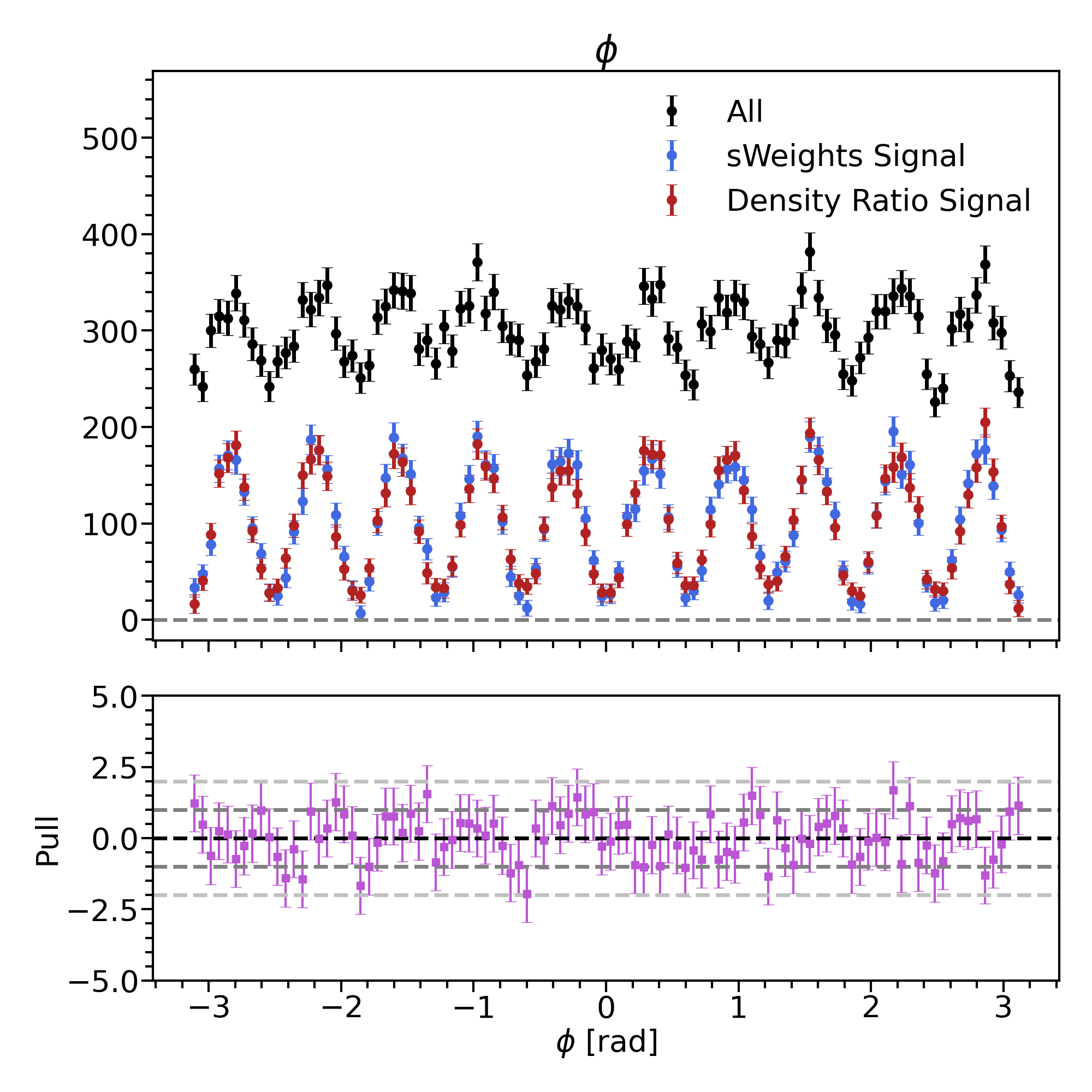}
    \includegraphics[width=0.49\textwidth]{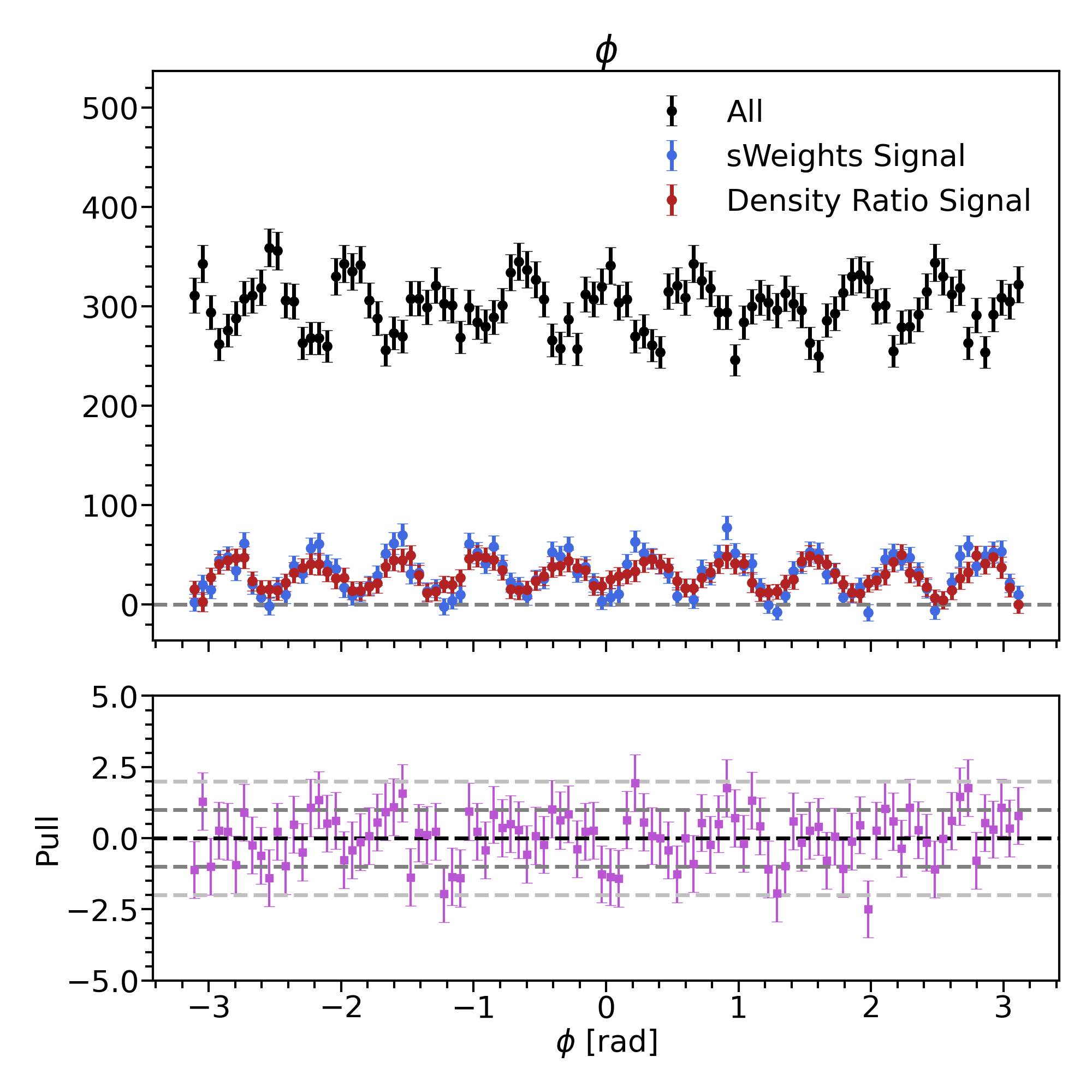}
   \caption[Higher Frequency distributions]{
   A comparison of all events, \emph{sWeights} signal events (blue) and \emph{drWeights} red for a cosine of frequency = 10, a for 1:2 (left) and 1:9 (right) signal-to-background ratio.}
    \label{fig:HighFrquencyPlots}
\end{figure}

\noindent
In Figure \ref{fig:HighFrquencyPlots} we show the full event, \emph{sWeight} and \emph{drWeight} distributions for a cosine of frequency 10, for the two different signal-to-background ratios we used previously. Visually, the \emph{sWeight} and \emph{drWeight} distributions agree well and the pull distributions are still within expected limits. On closer inspection, it is evident that the larger background sample has slightly larger peaks and dips for the blue \emph{sWeight} distribution than the red \emph{drWeight} one. This leads to the \emph{drWeight} events having a smaller fitted amplitude than for the \emph{sWeights}.\\

\begin{figure}[ht!]
    \centering   
    \includegraphics[width=0.9\textwidth]{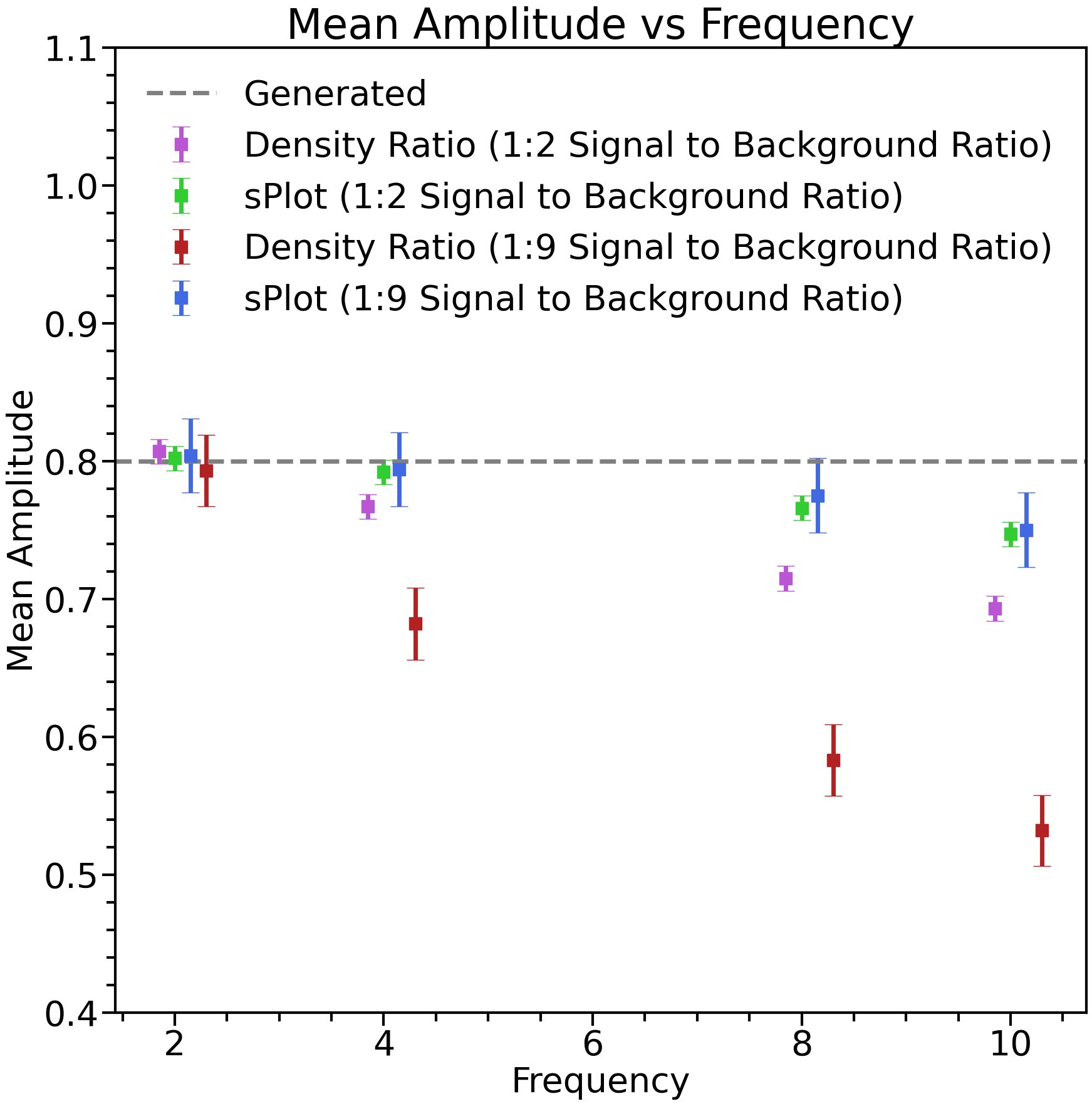}
     \caption[Higher Frequency results]{
     Effect on the fitted cosine amplitude from increasing its frequency. We compare the \emph{sWeight} and \emph{drWeight} cases for different signal-to-background ratios.
   }
    \label{fig:HighFrquencyResults}
\end{figure}

\noindent
In Figure~\ref{fig:HighFrquencyResults} we systematically increase the frequency from 2 to 10 and report the fitted amplitude averaged over 50 fits, for the two types of weights and two signal-to-background ratios.
We observe a steady fall in the \emph{drWeights} amplitudes that is significantly stronger for the larger background sample. This highlights the effect of the smoothing in the algorithm; it has a similar effect to adding experimental smearing. It is therefore important when deciding to use this method to evaluate if such limitations will significantly affect the results in a particular use-case.\\

\begin{figure}[ht!]
    \centering   
    \includegraphics[width=0.9\textwidth]{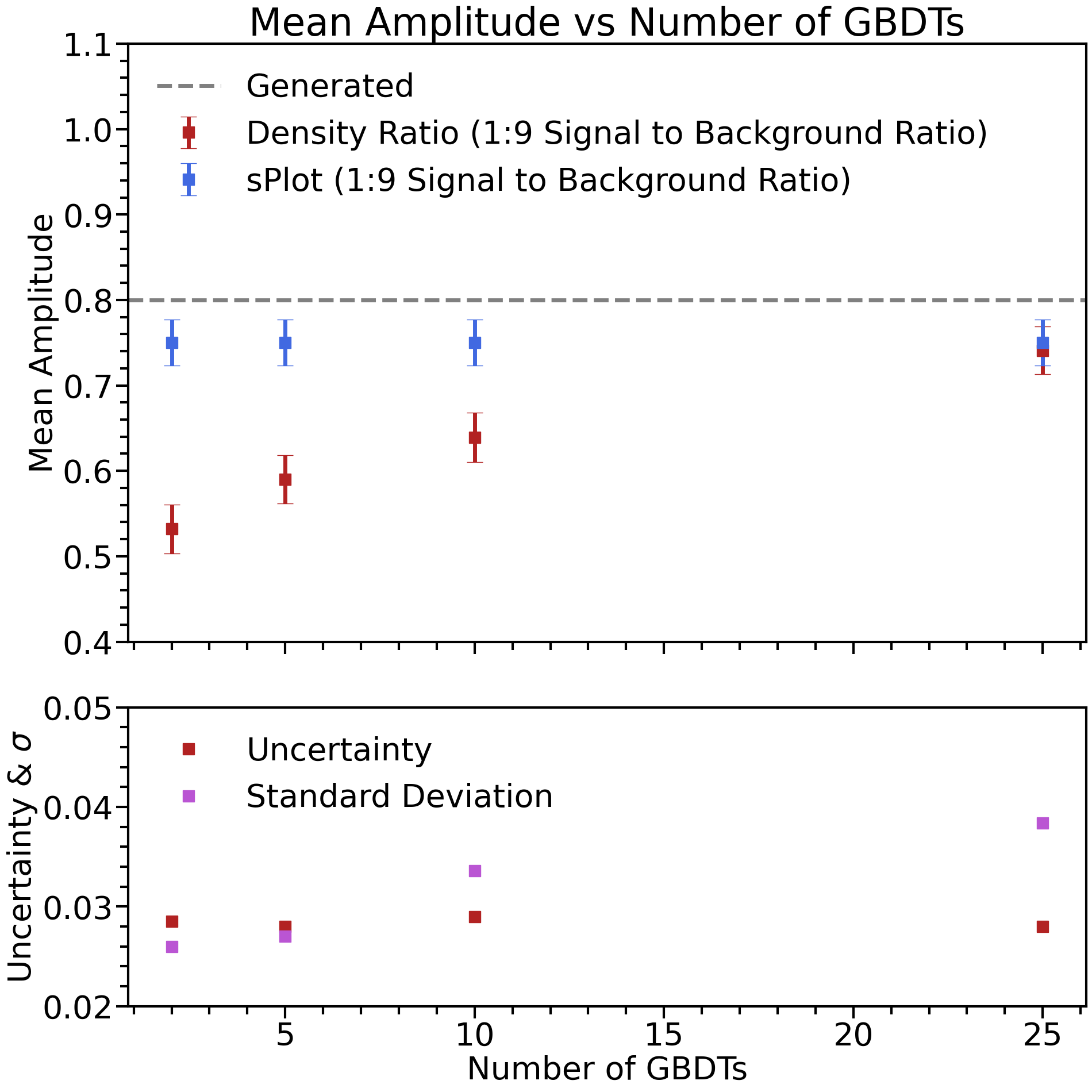}
     \caption[Higher Number of converters.]{
     Effect on the fitted cosine amplitude from increasing the number of GBDTs, for a cosine frequency of 10. We compare the \emph{sWeight} and \emph{drWeight} cases and the uncertainty and standard deviations of the \emph{drWeights}.
   }
    \label{fig:HighNGBDTsResults}
\end{figure}

\noindent
As shown in table~\ref{tab:toy_models}, the density ratio product performs consistently better than a single density ratio. This is due to the second reweighting step correcting the bias introduced by the smoothing of the first model. We then increased the number of converters from 2 to 25. Figure~\ref{fig:HighNGBDTsResults} shows the impact on the \emph{drWeights} amplitudes, along with a comparison between the uncertainty and the standard deviation, for a cosine frequency of 10. Adding further converters allows to correct the amplitude up until the \emph{drWeights} reproduce the \emph{sWeights} amplitude. However the standard deviation becomes larger than the uncertainty, pointing to the fact that there is larger than expected fluctuation in results between training iterations. From a practical perspective this should not be an issue when using the \emph{drWeights} method, for example to create datasets from experimental for the purposes of training machine learning algorithms. The training of the \emph{drWeights} converters can be repeated many times with the best iteration used in further applications. Training fewer density ratio converters will produce stabler results, and decreases the training and prediction times. A single density ratio product should therefore be preferred over many density ratios, but this may be a viable solution in complicated scenarios as shown here.\\

\noindent
The improvement in the \emph{drWeights} amplitude is due to the subsequent reweighting steps correcting the smearing introduced by previous steps which smoothed over features in the signal distribution. However, it must be understood that adding many more reweighters injects further correlations between events through the estimation of the local density ratio, to better reproduce the \emph{sWeighted} distributions and reduce bias. In noisy data, this procedure can lead to more stable results, as we effectively smooth out the noise if the induced bias is less than the noise effects. An example of this is shown in Table~\ref{tab:toy_nEvs} in Appendix~\ref{app_tests}, for the low event-number tests. In contrast, \emph{sWeights} are model agnostic and simply provide the best unbiased estimate of the control variables given the a priori knowledge available for the discriminating variables. The \emph{drWeights} are therefore a useful tool to reproduce the \emph{sWeighted} distributions, but they are by construction biased.\\
 
\noindent
The fact that the standard deviation becomes larger than the uncertainty when using many \emph{drWeights} converters also points to some limitations in the statistical properties of \emph{drWeights}, especially in scenarios such as the one described above with sharp features and large backgrounds.Table~\ref{tab:toy_models} showed that the uncertainty in the data sets produced with \emph{drWeights}, which therefore have correlated bins due to the smoothing and bias described above, is generally somewhat larger than that obtained from an uncorrelated dataset, for example with \emph{sWeights}. In Section~\ref{sect_methodDR} we suggested carrying over the uncertainty from the \emph{sWeights} uncertainty calculation; our results show that this may, however, underestimate the uncertainty on the \emph{drWeights} distributions, perhaps due to a lack of precision inherent in the training procedure. Therefore,  in general, \emph{sWeights} should be preferred in cases where their well defined statistical properties and uncertainties are required. In turn, \emph{drWeights} should be used for practical applications where positive definite probabilities are required, such as creating machine learning training samples from experimental data containing contributions from several different event sources.\\

\subsection{$ep \to e' \pi^{+} n$  in CLAS12} \label{sect_casestudies_c12}
\noindent
This section will demonstrate an application of the method presented in Section~\ref{sect_method} to experimental nuclear physics data. The Continuous Electron Beam Accelerator Facility (CEBAF)~\cite{CEBAF2} delivers an electron beam with energy up to 12 GeV to the four experimental halls at the Thomas Jefferson National Accelerator Facility (JLab). The CEBAF Large Acceptance Spectrometer at 12 GeV (CLAS12)~\cite{C12Overview} is located in Hall B. The CLAS12 experimental program broadly encompasses electroproduction experiments aiming to further the global understanding of hadronic structure and Quantum Chromodynamics~\cite{C12Program}. The CLAS12 detector was built to have full azimuthal angular coverage and a large acceptance in polar angle, allowing measurements to be made over large kinematic ranges~\cite{C12Overview}. Very low polar angular coverage, from 2.5 to 5 degrees, is enabled by the forward tagger (FT), whilst the forward detector (FD) covers the range of polar angles from 5 to 35 degrees and is segmented into six sectors of azimuthal angle. The central detector (CD) covers the polar angular range of 35 to 125 degrees.\\

\noindent
The Forward Electromagnetic Calorimeters (ECAL)~\cite{ECAL} are employed to detect and identify neutrons in the FD. Studies of neutron detection in the FD, for example measuring the neutron detection efficiency or establishing corrections to the measured neutron momenta, often employ the exclusive reaction $ep \to e' \pi^{+} n$. A neutron reconstructed from the reaction $ep \to e' \pi^{+} X$ is compared to the detected neutron, allowing estimates of quantities like the neutron detection efficiency. Several analysis procedures are made to check that the detected neutron corresponds to the reconstructed neutron, such as restricting the direction of the reconstructed neutron to the fiducial region of the FD and requiring that the reconstructed and detected neutron have a small difference in polar and azimuthal angles.\\

\begin{figure}[ht!]
    \centering   
    \includegraphics[scale=0.14]{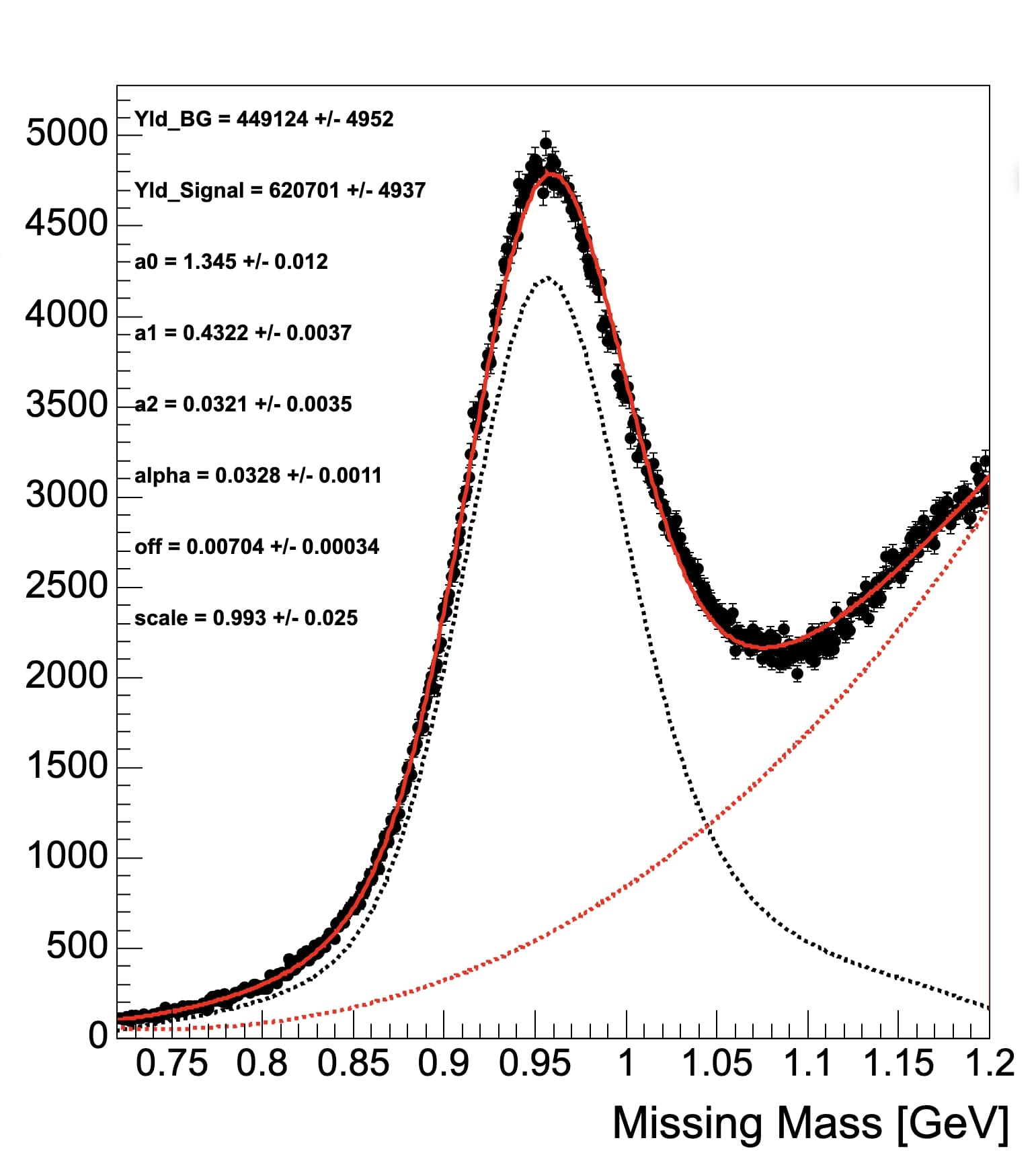}
    \caption[Fitting the neutron missing mass with \emph{sPlot}.]{An \emph{sPlot} fit of the invariant mass of $ep \to e' \pi^{+} X$ in the region of the neutron missing mass. The background is fitted with a third order polynomial function. The parameters Yld\_BG relate to the background yield, with a1, a2 and a3 the polynomial coefficients. The neutron signal is fitted using a simulated template. The parameter Yld\_Signal relates to the neutron signal yield. Three parameters, alpha, off, scale, allow to convolute the simulated model to represent the data. The parameter off allows to offset the mean of the distribution, with the scale parameter scaling the width. The parameter alpha allows to convolve the simulated model with an additional Gaussian to better fit the data.}
    \label{fig:MMFit}
\end{figure}

\noindent
In this analysis the mass of the missing particles in the reaction $ep \to e' \pi^{+} X$ close to the neutron mass was fitted to estimate the number of signal events where the exclusive reaction $ep \to e' \pi^{+} n$ was produced. Figure~\ref{fig:MMFit} shows the fit of the missing mass of $ep \to e' \pi^{+} X$ with the background described by a third order polynomial. The neutron signal was given by a template histogram from  simulated data that was created by generating events for the reaction $ep \to e' \pi^{+} n$ and running them through the CLAS12 simulation framework, GEMC~\cite{GEMC}. This fit allowed \emph{sWeights} to be assigned to separate the neutron signal from the underlying background. We are now able to disentangle neutron and background distributions. However if we wish to train a machine learning algorithm with these signal neutrons we should convert the \emph{sWeights} to \emph{drWeights} as described in Section~\ref{sect_method}, by training a density ratio converter with their three momentum components as input features.\\

\noindent
The methodology of Section~\ref{sect_method} was applied to the \emph{sWeights} produced by the fit in Figure~\ref{fig:MMFit}. We employed a density ratio product with two gradient boosted decision tree (GBDT) steps with a max depth of ten each implemented with the scikit-learn library~\cite{scikit-learn}. The denominator sample was composed of all unweighted events and the numerator sample was composed of all events weighted using the neutron signal \emph{sWeights}. The GBDTs were trained on the reconstructed neutron spherical momentum components.\\

\begin{figure}[ht!]
    \centering   
    \includegraphics[width=0.45\textwidth]{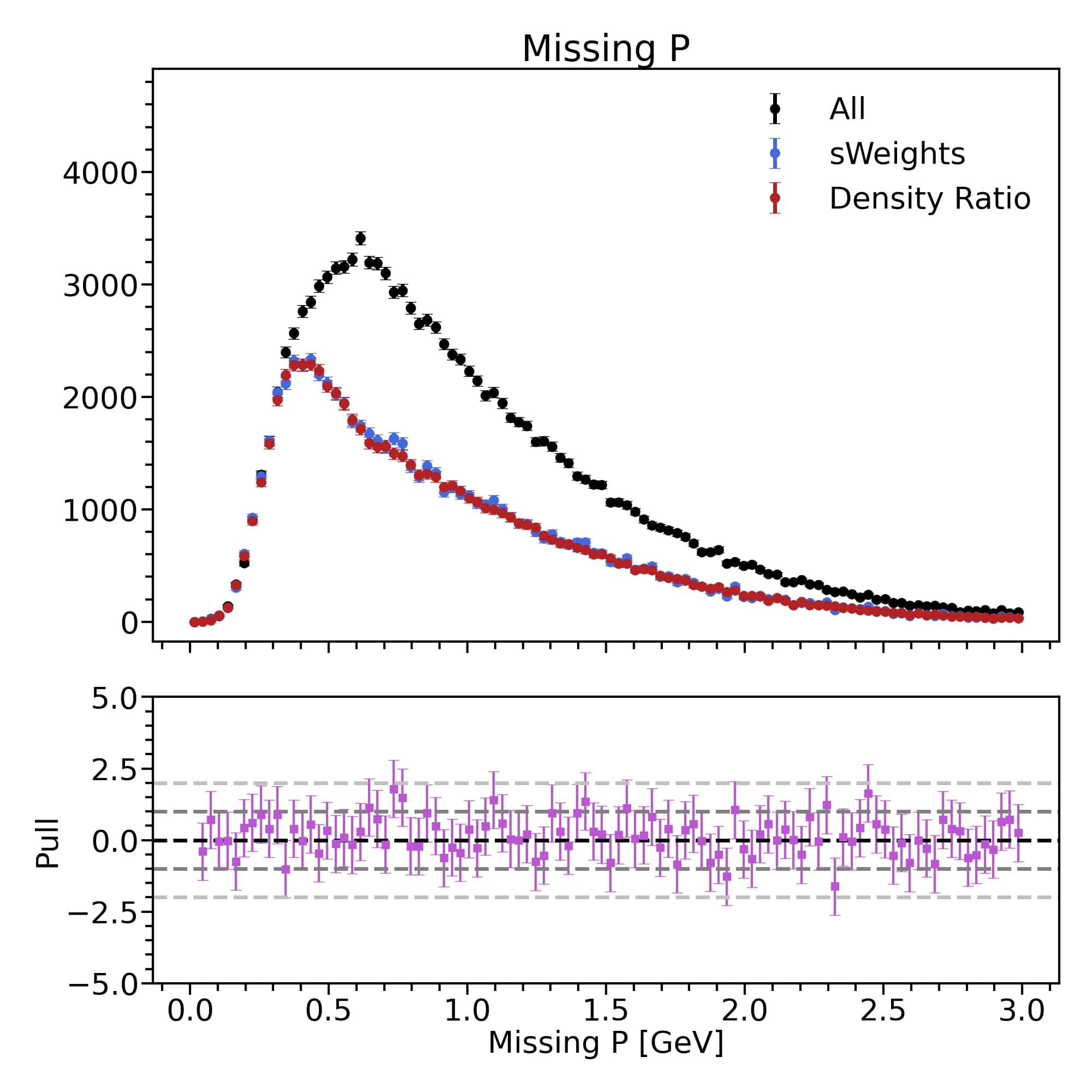}
    \includegraphics[width=0.45\textwidth]{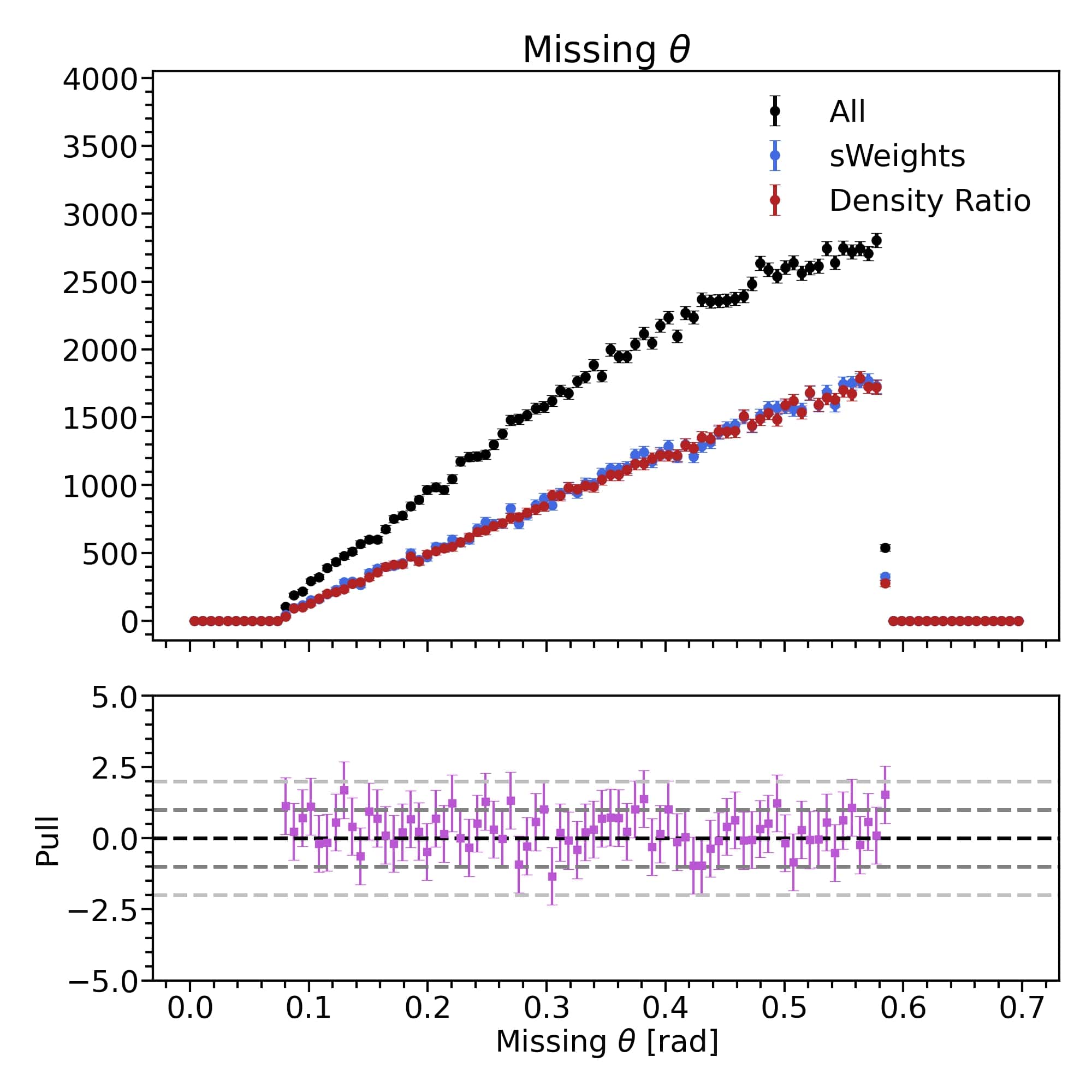}
    \includegraphics[width=0.45\textwidth]{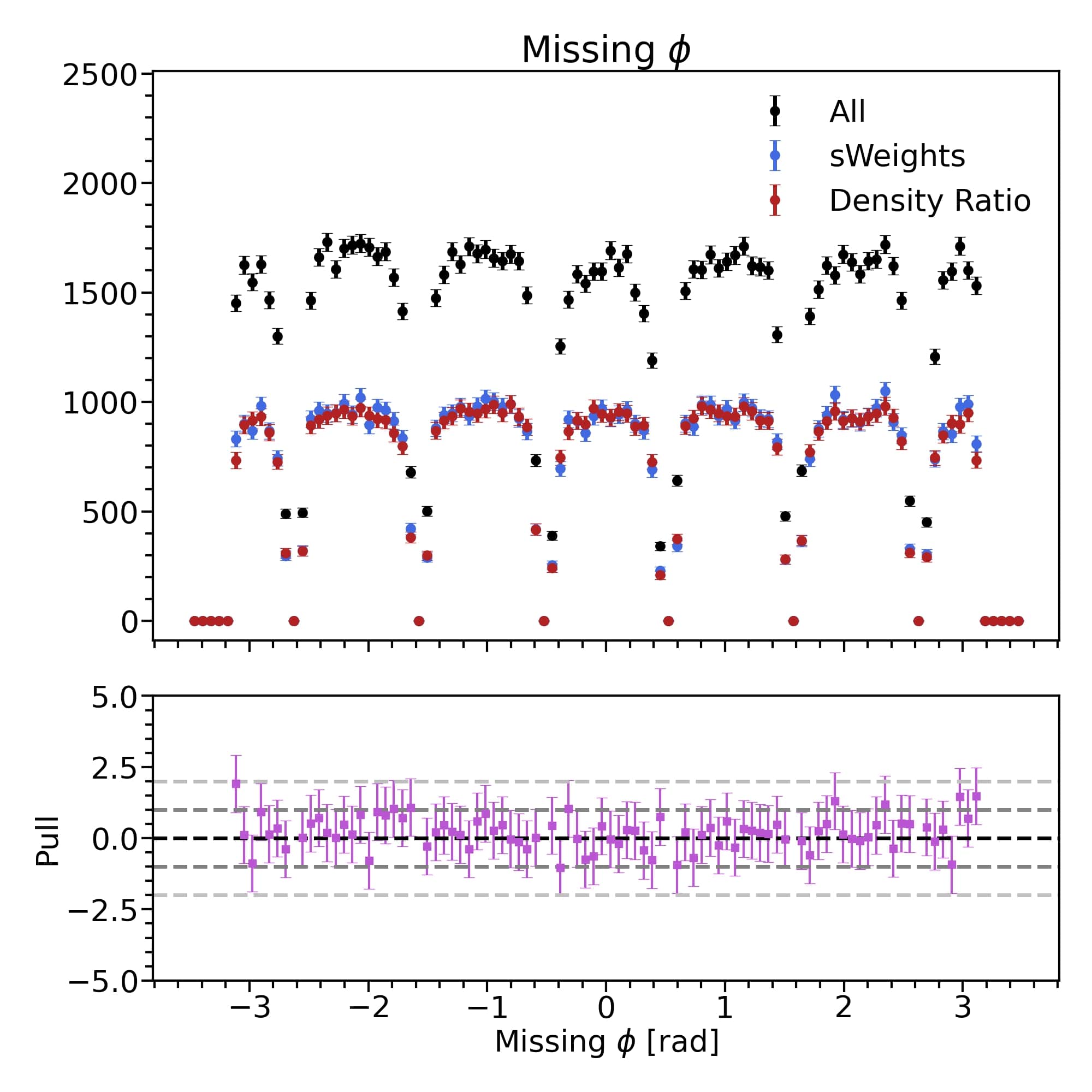}
   \caption[Comparing the \emph{drWeight} and \emph{sWeight} neutron signal in CLAS12 data.]{A comparison of all events (black) and all events weighted by the \emph{sWeights} (blue) or by the \emph{drWeights} (red) using the density ratio method. The comparison is made for each of the missing momentum (left), polar angle ($\theta$, right) and azimuthal angle ($\phi$, bottom).}
    \label{fig:NeutronResults1}
\end{figure}

\begin{figure}[ht!]
    \centering   
    \includegraphics[width=0.95\textwidth]{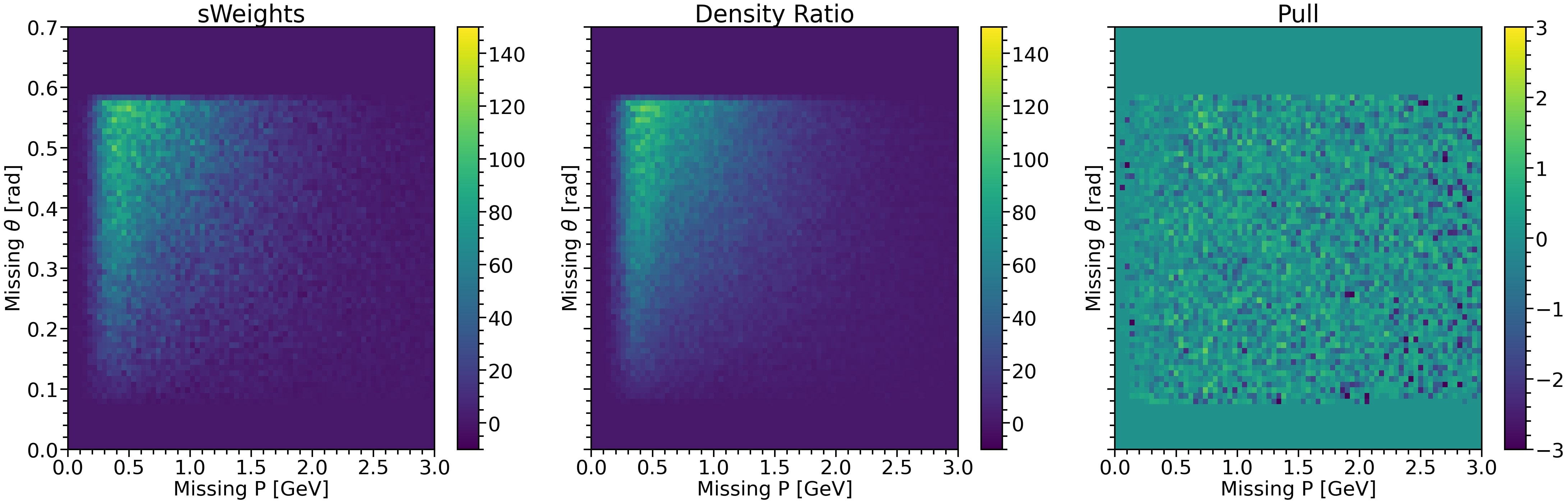}\\
    \includegraphics[width=0.95\textwidth]{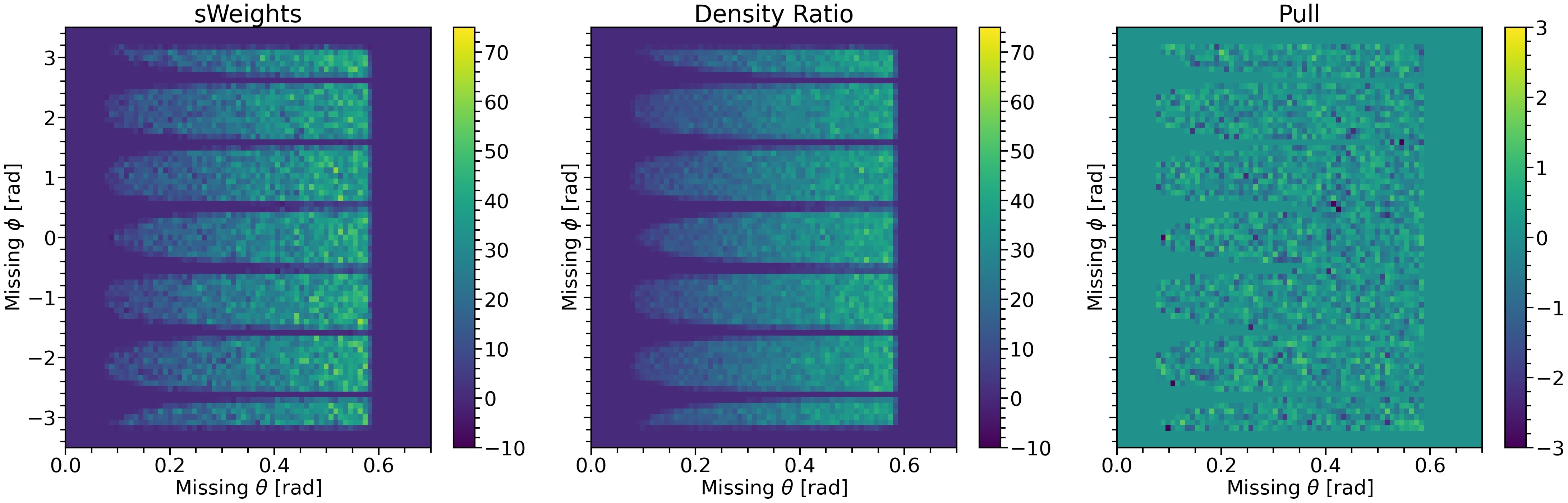}
    \caption[Comparing the \emph{drWeight} and \emph{sWeight} neutron signal in CLAS12 data.]{Top Row: The momentum and the polar angle ($\theta$) correlation histogram weighted using the \emph{sWeights} (left) or \emph{drWeights} (middle) and the pull between the two (right). Bottom Row: The polar angle ($\theta$) and azimuthal angle ($\phi$) correlation histogram weighted using the \emph{sWeights} (left) or \emph{drWeights} (middle) and the pull between the two (right).}
    \label{fig:NeutronResults2}
\end{figure}

\noindent
Figures~\ref{fig:NeutronResults1} and~\ref{fig:NeutronResults2} show a comparison of all events, and the same events weighted by the \emph{sWeights} and the \emph{drWeights}, for each of the reconstructed momentum magnitude, polar and azimuthal angles. The uncertainties were propagated from the original \emph{sWeights}. These figures also show the pull between the two distributions defined as the difference between the bin values for the two distributions divided by the sum squared of their uncertainties. The correlation between momentum and polar angle, and polar and azimuthal angles are also shown. Overall, the distributions are well reproduced when applying the \emph{drWeights}. Accurately converting the \emph{sWeights} to positive definite probabilities could then have many applications. The \emph{drWeights} could be used, for example, to train a machine learning algorithm to model the neutron detection efficiency as a function of all three momentum components reconstructed from the missing four-vector.\\

\section{Conclusion And Outlook} \label{sect_ccl}
\noindent
The \emph{sPlot} formalism is a useful tool to separate different event species in high energy and nuclear physics experimental data. However, the \emph{sWeights} obtained using the \emph{sPlot} technique have negative values which will lead to issues in training and poor overall performance when using machine learning algorithms on data weighted with these \emph{sWeights}. \\

\noindent
Ref.~\cite{NeuralResampler} introduced a methodology based on density ratio estimation to resample weights from Monte Carlo event generators so as to avoid negative weights. The key point is that setting up the training sample so as to learn the density ratio of the weights, which in turn allows to resample these, means that the binary cross entropy loss function is not detrimentally affected by negative weights (c.f. Equation~\ref{eqn:neuralresampler_loss}). This article has detailed how \emph{sWeights} can be converted to probabilities using density ratio estimation, producing \emph{drWeights}. We further detailed how gradient boosted decision trees are particularly well suited to converting \emph{sWeights}. Moreover density ratio products, where a subsequent reweighting is applied to fine-tune the distributions, produces particularly robust weights for the dataset. \\

\noindent
Three case studies were presented, one based on a simple toy example, the second extending this to more complicated situations and another on experimental data from the CLAS12 detector. These demonstrated the potential of density ratios to convert \emph{sWeights}. The first case study also highlighted the advantages of using decision trees, namely a non negligible increase in prediction rate and good performance with decreased data sizes and signal to background ratios. The latter is an important consideration as experimental data is limited and may have irreducible signal to background ratios. It is generally beneficial to have several options of learning algorithms to choose from depending on the dataset and task at hand.\\

\noindent
However, the first case study also demonstrated that the \emph{drWeights} converter introduces some degree of smoothing, or correlating of events, as the converter averages over local densities when producing the \emph{drWeights}. The second case study showed that when backgrounds are large and distribution features are sharp, these smoothing effects can lead to significant loss in fidelity. This can be compensated by introducing many subsequent reweighting steps at a loss in training stability and training and prediction times. As such, care should be taken in evaluating the applicability and implementation of the density ratio weight conversion to given use cases. Nevertheless, the \emph{drWeights} are generally a useful tool for converting \emph{sWeights} to probabilities.\\

\noindent
The final case study demonstrated how the density ratio conversion of \emph{sWeights} can be well suited to real experimental data. The methodology presented in this article can be used for many applications of machine learning to experimental data in high energy and nuclear physics. In general, it is necessary to disentangle the distributions of different event species in experimental data to be able to use the distributions of specific species to train a machine learning algorithm. For example, converting \emph{sWeights} to \emph{drWeights} as shown in the third case study could be applied as a useful first step to learning the neutron detection efficiency in the CLAS12 forward detector. The mass of missing particles in $ep \to e' \pi^{+} X$ can be fitted to distinguish the neutron signal from the background. The signal \emph{sWeights} can then be converted to \emph{drWeights} as described in Section~\ref{sect_casestudies_c12}. These converted weights can then be used in training a binary classification algorithm to model the CLAS12 forward detector neutron detection efficiency. Previous work has demonstrated the use of density ratio estimation via binary classification to model detector efficiency from simulation~\cite{OmniFold,macparticlesPaper}. The method to obtain~\emph{drWeights} described in this article could be a first step to modeling detector efficiency from experimental data. In general, many such applications will exist and require to accurately convert \emph{sWeights} to positive definite probabilities as done here.\\

\noindent
Finally we note that for general analysis, for example extracting an amplitude from a background subtracted distribution, \emph{sWeights} themselves will still provide a more reliable method. The goal here is not to replace \emph{sWeights} in these circumstances, only in the cases such as machine learning training where positive definite probabilities are required.\\

\section*{Acknowledgements}
\noindent
The authors thank Simon Gardner for reviewing the text. We would also like to thank the CLAS Collaboration for providing data used in this body of work.
This material is based upon work supported by the U.S. Department of Energy, Office of Science, Office of Nuclear Physics under contract DE-AC05-06OR23177 and by the U.K. Science and Technology Facilities Council under grant ST/V00106X/1.\\



\appendix

\section{Appendix: Further Tests} \label{app_tests}

\noindent
This appendix will present and describe the results of the tests mentioned in Section~\ref{sect_casestudies}. As a reminder, the mean amplitude and uncertainty are obtained from 50 iterations of generating a toy dataset, producing the \emph{sWeights}, training and applying the \emph{drWeights}. The expectation is that the mean should be consistent with the nominal value of 0.8 used to generate the signal $\phi$ distribution, whilst the mean uncertainty ($\Bar{\sigma}_{fit}$) and standard deviation on the amplitude ( $\hat{\sigma}_{rms}$) should be consistent. The first row of each table will show the results for the sWeighted distributions that we are trying to emulate.\\

\begin{table}[ht!]
\centering
\begin{tabular}{ |c||c|c|c|c| } 
\hline
GBT & & &  &  \\
Depth & Mean & $\hat{\sigma}_{rms}$ & $\Bar{\sigma}_{fit}$ & $\frac{ \hat{\sigma}_{rms}}{\Bar{\sigma}_{fit}}$ \\
\hline
\emph{sWeights}& & &  & \\
 & 0.802 & 0.0082 & 0.0089 & 0.92  \\
\hline
\emph{drWeights} & & &  &  \\
 Depth 3 & 0.770 & 0.0105 & 0.0090 & 1.17  \\
\hline
\emph{drWeights}& & &  & \\
Depth 5 & 0.777 & 0.0103 & 0.0090 & 1.14  \\
\hline
\emph{drWeights}& & &  & \\
Depth 10 & 0.810 & 0.011 & 0.0092 & 1.20  \\
\hline
\emph{drWeights}& & &  & \\
Depth 25 & 0.834 & 0.0112 & 0.0103 & 1.10  \\
\hline
\emph{drWeights}& & &  & \\
Depth 50 & 0.786 & 0.0090 & 0.0096 & 0.94  \\
\hline
\end{tabular}\\

  \caption[Comparing the performance of density ratio GBDTs with various depth.]{Comparison of the $\phi$ amplitude measured for GBDTs with various depth used by the density ratio estimation. The signal distribution was generated with a $\phi$ amplitude of 0.8 for $10^5$ events with a signal to background ratio of 1:2. The data generation and training were repeated 50 times. The mean and standard deviation ($\sigma$) of the measured amplitudes are reported, along with the mean uncertainty.}
    \label{tab:toy_depth}
\end{table}

\noindent
Table~\ref{tab:toy_depth} shows the performance of a single GBDT converter at various depths. Shallower GBDTs seem to perform less well; this is most likely due to their inability to fully capture correlations in the input variable space. This conclusion was also reached when using the density ratio method for fast detector acceptance simulations~\cite{macparticlesPaper}. At a depth of 10 the GBDT performs adequately. The training and prediction rates of a GBDT will generally increase with its depth, and so we chose to limit the depth at 10.\\

\begin{table}[ht!]
\centering
\begin{tabular}{ |c||c|c|c|c| } 
\hline
Model & & &  &  \\

Signal:Background & Mean & $\hat{\sigma}_{rms}$ & $\Bar{\sigma}_{fit}$ & $\frac{ \hat{\sigma}_{rms}}{\Bar{\sigma}_{fit}}$ \\
\hline
\emph{sWeights}& & &  & \\
1:2 & 0.802 & 0.0082 & 0.0089 & 0.92  \\
1:9 & 0.804 &  0.0244 & 0.0274 & 0.89 \\
\hline
GBDT & & &  &  \\
 1:2 & 0.796 & 0.0094 & 0.0091 & 1.03  \\
 1:9 & 0.707 & 0.0779 & 0.0271 & 2.87 \\
\hline
GBDT \& GBDT& & &  & \\
1:2 & 0.807 & 0.0093 & 0.0092 & 1.01  \\
1:9 & 0.793 & 0.0260 & 0.0285 & 0.91 \\
\hline
GBDT \& HistGBDT& & &  & \\
1:2 & 0.810 & 0.011 & 0.0092 & 1.20  \\
1:9 & 0.791 & 0.0347 & 0.0283 & 1.23 \\
\hline
HistGBDT & & &  &  \\
1:2 & 0.760 & 0.0169 & 0.0089 & 1.90  \\
1:9 & 0.643 & 0.0356 & 0.0261 & 1.36 \\
\hline
HistGBDT \& GBDT & & &  & \\
1:2  & 0.788 & 0.0115 & 0.0091 & 1.26  \\
1:9 & 0.739 & 0.0326 & 0.0281 & 1.16 \\
\hline
HistGBDT \& HistGBDT & & &  &  \\
1:2  & 0.782 & 0.0112 & 0.0090 & 1.24  \\
1:9 & 0.713 & 0.0342 & 0.0274 & 1.25\\
\hline
NN & & &  &  \\
1:2  & 0.782 & 0.0262 & 0.0089 & 2.94  \\
1:9 & 0.743 & 0.0417 & 0.0264 & 1.58 \\
\hline
NN \& GBDT& & &  &  \\
1:2 & 0.822 & 0.0130 & 0.0093 & 1.40 \\
1:9 & 0.826 & 0.0369 & 0.0299 & 1.23 \\
\hline
NN \& HistGBDT & & &  & \\
 1:2  & 0.813 & 0.0180 & 0.0093 & 1.94  \\
 1:9 & 0.816 & 0.0351 & 0.0291 & 1.21\\
\hline
\end{tabular}\\

  \caption[Comparing the performance of density ratio models for various models.]{Comparison of the $\phi$ amplitude measured for various models used by the density ratio estimation. The signal distribution was generated with a $\phi$ amplitude of 0.8 for $10^5$ events with a signal to background ratio of 1:2 or a signal to background ratio of 1:9. The data generation and training were repeated 50 times. The mean and standard deviation ($\sigma$) of the measured amplitudes are reported, along with the mean uncertainty.}
    \label{tab:toy_models}
\end{table}

\noindent
In Table~\ref{tab:toy_models} we show the results of the different converters. We observe that all except the HistGBDT perform adequately in reproducing the amplitude (mean value) for the 1:2 signal to background case. In particular, models with a density ratio product, or reweighting step, are an improvement on the single-step case. For the 1:9 case where the background dominates, models with the GBDT reweighting step perform well, while the others do not give an as accurate amplitude. In all cases the uncertainty is consistent as it is just a property of the data statistics and we are using the sWeighted sum of the weights squared. The standard deviation, however, does vary significantly from the given uncertainty showing a lack of robustness in the procedure for some converter models. This is likely due to some random events getting inaccurate weights and thereby less accurate $\phi$ distributions. This also seems to be more of an issue for the neural network based models and in the higher background tests. On the other hand, the double GBDT model seems to perform admirably even with 1:9 signal to background and would seem to be the most robust converter.\\

\noindent
The number of generated events was then varied to ascertain the impact this has on the density ratio estimation using the preferred GBDT and GBDT density ratio product. The results are shown in Table~\ref{tab:toy_nEvs}, together with the results from the \emph{sWeights} fitting. The amplitude was again generated at 0.8, and here the signal to background ratio was fixed to the large background case at 1:9. In all cases, even with as few as $10^3$ generated events, the amplitude measured with the \emph{drWeights} is consistent with the generated amplitude of 0.8. The $\Bar{\sigma}_{fit}$ and $\hat{\sigma}_{rms}$ from the amplitude fits are generally consistent. At 1000 events, corresponding to 100 actual signal events, $\Bar{\sigma}_{fit}$ is twice the size of $\hat{\sigma}_{rms}$, suggesting the uncertainty is overestimated by a factor 2. This is not surprising given that the sum of the squared weights contribution to the uncertainty is less valid at low statistics. Also at 1000 events the \emph{drWeights} clearly outperform the \emph{sWeights}. This is due to the latter producing bins with unphysical negative counts, a problem which is resolved by using probability weights. This issue is ultimately an artifact of the weighted binned fitting, which is not the ideal method for extracting parameters from data. Instead, one should use an event-by-event maximum likelihood procedure to produce reliable results, but this is outside the scope of this work.\\

\begin{table}[ht!]


\centering
\begin{tabular}{ |c||c|c|c|c|} 
\hline
\# Events  & & & & \\ 
Weights & Mean & $\hat{\sigma}_{rms}$ & $\Bar{\sigma}_{fit}$ & $\frac{ \hat{\sigma}_{rms}}{\Bar{\sigma}_{fit}}$ \\
& & &  &   \\
\hline
$10^3$ & & &  &   \\
\emph{sWeights} & 17.94 & 84.81 & 14.67 & 5.78 \\
\emph{drWeights}  & 0.679 & 0.2710 & 0.5902 & 0.46\\
\hline
$10^4$& & &  &   \\
\emph{sWeights} & 0.870 & 0.1090 & 0.0953 & 1.14  \\
\emph{drWeights}  & 0.778 & 0.0929 & 0.1038 & 0.89 \\
\hline
$10^5$ & & &  &   \\
\emph{sWeights} & 0.804 & 0.0244 & 0.0274 & 0.89  \\
\emph{drWeights}   & 0.793 & 0.0260 & 0.0285 & 0.91 \\
\hline
$10^6$ & & &  &  \\
\emph{sWeights} & 0.799 & 0.0104 & 0.0090 & 1.16 \\
\emph{drWeights} & 0.792 & 0.0110 & 0.0092 & 1.20  \\
\hline
\end{tabular}\\

  \caption[Comparing the performance of density ratio models as a function of the number of generated events.]{Comparison of the $\phi$ amplitude measured on the \emph{sWeights} signal distribution and the \emph{drWeights} (DR) signal distribution when increasing the number of generated events. The signal distribution was generated with a $\phi$ amplitude of 0.8 and a signal to background ratio of 1 to 9. The data generation and training were repeated 50 times. The mean and standard deviation ($\hat{\sigma}_{rms}$) of the measured amplitudes are reported, along with the mean uncertainty from the fit, $\Bar{\sigma}_{fit}$. The \emph{drWeights} are estimated using the GBDT and GBDT density ratio product. \textbf{Note} that to extract the $\phi$ amplitude, $\phi$ was binned in 100 bins for all tests except at $10^3$ events where $\phi$ was binned in 50 bins due to lower statistics.}
    \label{tab:toy_nEvs}
\end{table}

\end{document}